\documentclass[12pt]{article}
\usepackage{setspace}
\usepackage[a4paper, total={6in, 8in}]{geometry}
\usepackage[spanish,english]{babel}
\usepackage{relsize}\usepackage{xcolor}
\usepackage{amsfonts,multirow,amsmath,adjustbox,natbib,bbm, bm}
\usepackage{accents}\usepackage{booktabs}\usepackage{lscape}
\usepackage{animate}
\usepackage{graphicx}
\usepackage{media9}
\usepackage{tikz}
\usepackage[utf8]{inputenc}
\usepackage{subcaption}
\usepackage{caption} 
\usepackage{longtable}
\usepackage{bm,bigints}
\usepackage{natbib}
\usepackage{dsfont}
\usepackage{graphicx}
\usepackage{wasysym} % Cents
\usepackage[textsize=tiny]{todonotes}
\usepackage{algorithm}

\usepackage{setspace} 
\makeatletter

\AtBeginDocument{
	\g@addto@macro{\appendix}{\renewcommand{\p@subsection}{}}
}
\makeatother
\usepackage{color,soul}

\usepackage{soul,color}
\usepackage[para,online,flushleft]{threeparttable}\usepackage{multirow}\usepackage{multicol}%\usepackage{subfigure}
\usepackage{graphicx, comment}
\definecolor{ahjcolor}{rgb}{0.0, 0.13, 0.40}			
\usepackage[colorlinks	=true,										% HYPERREFERENCE
linkcolor	=ahjcolor,									%
urlcolor	=ahjcolor,									%
citecolor	=ahjcolor,
bookmarksnumbered=true]{hyperref}

\usepackage{algorithm}

\makeatletter
\def\BState{\State\hskip-\ALG@thistlm}
\makeatother
\def\1{1\!{\rm l}}

\newlength\figureheight
\newlength\figurewidth

\usepackage{enumitem,float,rotating,multirow,algorithm,algpseudocode,graphicx,lscape,tabularx,ragged2e}

\usepackage{graphicx,bm}

\providecommand{\Keywords}[1]
{
	\small	
	\textbf{\textit{Keywords:}} #1
}

\providecommand{\JEL}[1]
{
	\small	
	\textbf{\textit{JEL:}} #1
}

\usepackage{footnote}
\makesavenoteenv{algorithm}

\title{Marijuana on Main Streets? The Story Continues in Colombia\\
An Endogenous Three-part Model\thanks{We acknowledge the supercomputing resources that were made available by the Centro de Computación Científica Apolo at Universidad EAFIT (http://www.eafit.edu.co/apolo) to conduct the research reported in this scientific product. All errors are our own.}
}
\author{Andr\'es Ramírez--Hassan\thanks{OMEGA research group, Universidad EAFIT, Medellín, Colombia. \href{aramir21@eafit.edu.co}{E-mail: aramir21@eafit.edu.co}. Carrera 49 7 Sur-50, phone: + 57 2619500.}, Catalina G\'omez-Toro\thanks{OMEGA research group, Universidad EAFIT, Medellín, Colombia. \href{cgomezt1@eafit.edu.co}{E-mail: cgomezt1@eafit.edu.co}}\\
Katherin Tangarife-Londoño\thanks{OMEGA research group, Universidad EAFIT, Medellín, Colombia. \href{atangarifel@eafit.edu.co}{E-mail: ktangarifel@eafit.edu.co}.}, Santiago Vel\'asquez\thanks{OMEGA research group, Universidad EAFIT, Medellín, Colombia. \href{svelas45@eafit.edu.co}{E-mail: svelas45@eafit.edu.co}.}}
\begin{document}
	\maketitle
	\begin{abstract}
Cannabis is the most common illicit drug, and understanding its demand is relevant to analyze the potential implications of its legalization. This paper proposes an endogenous three-part model taking into account incidental truncation and access restrictions to study demand for marijuana in Colombia, and analyze the potential effects of its legalization. Our application suggests that modeling simultaneously access, intensive and extensive margin is relevant, and that selection into access is important for the intensive margin. We find that younger men that have consumed alcohol and cigarettes, living in a neighborhood with drug suppliers, and friends that consume marijuana face higher probability of having access and using this drug. In addition, we find that marijuana is an inelastic good (-0.45 elasticity). Our results are robust to different specifications and definitions. If marijuana were legalized, younger individuals with a medium or low risk perception about marijuana use would increase the probability of use in 3.8 percentage points, from 13.6\% to 17.4\%. Overall, legalization would increase the probability of consumption in 0.7 p.p. (2.3\% to 3.0\%). Different price settings suggest that annual tax revenues fluctuate between USD 11.0 million and USD 54.2 million, a potential benchmark is USD 32 million.    
		
	\end{abstract}
	\JEL{D12, H25, K14, K42}
	
	\Keywords{Marijuana demand, Marijuana legalization, Three-part model, Truncation.}
	
	\clearpage
	
	\maketitle

 \section{Introduction}\label{sec:Int}
Understanding the demand for illicit drugs is relevant, as it is a necessary analysis to comprehend the potential effects of their legalization. Cannabis remains the most commonly used illicit drug worldwide, with an estimated 209 million consumers in 2020, reflecting a 23\% increase over the last decade. This is followed by opioids (61 million), amphetamines (34 million), cocaine (21 million), and ecstasy (20 million) \citep{UNODC2022}. For decades, several countries have entered the debate on the legalization of marijuana. In this paper, we provide a new methodology for examining the consequences of legalizing marijuana, which helps to answer questions regarding its access, and extensive and intensive margins. Therefore, we can have a better understanding on: how much the prevalence and intensity of marijuana use rise under legalization?, is there heterogeneity in response to legalization among different age groups?, and could government policies based on taxation and campaigns about risk perception be effective in curbing use? 

We extend \cite{jacobi2016marijuana}'s proposal, who proposed two independent two-part models to study the extensive and intensive margins of marijuana demand taken into account access restrictions when modeling demand for illicit drugs in Australia. In particular, we propose an endogenous three-part model that simultaneously takes into account access, extensive and intensive margins, as well as incidental truncation. To the best of our knowledge, this is the first paper to estimate demand for an illicit drug that considers simultaneously these features. Taking into account access restrictions is relevant as illicit drugs are not as easy to find, thus non-users have little information about how to get marijuana, which is a necessary condition to becoming a user \citep{jacobi2016marijuana}. Truncation is also relevant when modeling demand for illicit drugs as individuals tend to withhold information about their use \citep{lloyd2013stigmatization}. We apply our methodological proposal to model the demand for marijuana in Colombia, and study the potential implications of the legalization of marijuana in this market, where marijuana is decriminalized, making different counterfactual exercises trying to respond relevant inquiries regarding marijuana legalization. Colombian market is significant as this is one of the world's main producers of marijuana \citep{UNODC2021}, resulting in relatively easy access and low prices. 

For decades, countries have faced pressures regarding the decision to decriminalize or legalize the marijuana market. It is important to note that countries can implement a policy of liberalizing the marijuana market through either full legalization or decriminalization. According to \cite{nkansah2016gateway}, legalization occurs when authorities approve the use of a substance previously prohibited by law, thereby eliminating the risk of arrest or fines. Decriminalization suspends criminal sanctions for using or possessing a particular substance; however, it maintains the substance's illegal status, allowing for potential punishment through civil fines, education, social work, and other measures. Uruguay was the first country to legalize recreational marijuana in 2013; some US states have also done so, as well as Canada in 2018 \citep{jacobi2016marijuana}.

The literature highlights that the use of illicit drugs entails high costs for society, such as pressure on health systems, productivity loss \citep{van2006cannabis}, delinquency, violence \citep{norstrom2014cannabis}, incarceration and costs of the criminal justice system, educational performance, child abuse, and corruption, among others \citep{maccoun1996assessing, wen2014effect}. There is also evidence that fewer young people consider cannabis consumption dangerous, which leads to the normalization of consumption behavior \citep{jarvinen2011normalisation}. Additionally, there is the hypothesis that the legalization or decriminalization of marijuana serves as an incentive for consumption for younger populations since the early onset of cannabis use substantially increases subsequent consumption rates \citep{pudney2004keeping}. This scenario implies that it is not easy for countries to decide to liberalize the market. Studies that show post-legalization effects, such as those of \cite{Rubin-Kahana2022} and \cite{roffman2016}, suggest that some negative repercussions may not manifest until 5 or 10 years later, and the same applies to positive impacts.

Those in favor of marijuana legalization argue for a reduction in violence generated by the illegal market \citep{donohue2010rethinking}, as well as a decrease in the cost of law enforcement (police and judiciary), a reduction in arrests \citep{roffman2016}. \cite{Mace2020, Irvine2020, jacobi2016marijuana, Caputo1994}, also speak about tax revenue that could benefit from legalization for the United States, Canada, and Australia, with funds allocated to education, sports, and other addiction programs. On the issue of taxation, \cite{miron2005budgetary} also provides insights. Those who oppose legalization refer to the negative aspects of legalization, such as the possibility of a price decrease due to the elimination of transaction costs associated with illegality \citep{becker2006market}. Another relevant aspect is the potency of cannabis and its psychoactive component, which can have increasingly harmful consequences for health, including mental health problems \citep{elsohly2016changes, van2006cannabis}.

Thus, there is evidence of increased hospitalizations and intoxications due to excessive marijuana consumption post-legalization \citep{roffman2016, Rubin-Kahana2022}, as well as psychotic and mental health effects \citep{Moran2022}. Studies show an increase in consumption due to legalization in the United States. The most recent ones, such as that of \cite{Mennis2023}, show that after the legalization of recreational marijuana, there was an increase in consumption, especially among teenagers. Additionally, researchers have found an increased frequency of marijuana use and a higher prevalence of symptoms associated with marijuana use disorder \citep{Kilmer2022}. \cite{Barker2021} found that legalization had more significant direct effects on those who were already marijuana consumers, surrounded by an increasingly favorable climate for consumption.

The Canadian case \citep{Rotermann2020} shows similar results, where the increase in consumption is mainly associated with men over 25 years old. However, it is essential to note that individuals between 15 and 17 years old are experiencing a decrease. At the same time, the prevalence of consumption remains stable, and there is a decrease in the acquisition of marijuana from illegal sources. \cite{Rubin-Kahana2022} found different sources with conflicting data for Canadian teenagers, some studies report an increase, but most do not show a pronounced increase, demonstrating ambiguity in the analysis. In the case of Uruguay, \cite{Laqueur2020} sought to study the effect of legalization on high school students in Montevideo and regions within the country after legalization, but found no evidence of an impact on cannabis consumption or perceived consumption risk. However, they did find an increase in student perception of cannabis availability after legalization.

In addition to the above, polydrug use, which involves the combination of several substances, has become more visible. Regarding this, the most common combination in the Americas is that of cannabis with stimulants (such as cocaine and ecstasy), followed by opioids with stimulants, and finally, cannabis with opioids. The growing trend of polydrug use poses significant risks to consumers due to the interaction between substances \citep{UNODC2022}, constituting an essential topic on the global public agenda. Furthermore, since the 1970s, the question has arisen as to whether access to marijuana leads to an increase in the use of other more problematic substances, with the so-called ``gateway hypothesis" \citep{DeSimone1998, Kandel1975}. Regarding this, \cite{Kandel1992} argue that the progression to more toxic drugs depends on prior use of cigarettes or alcohol, then marijuana, and then more harmful substances. Among the new trends in consumption are an expansion in the forms of use and more potent products \citep{Hammond2021, Rubin-Kahana2022}. Furthermore, researchers \citep{Guttmannova2021, Kim2021} discovered a positive association between the frequency of cannabis and alcohol consumption. Additionally, \citep{Weinberger2022} established a positive relationship between marijuana consumption and frequency among tobacco users.

Finally, different studies have attempted to estimate the price elasticity of marijuana demand in the US, finding that it is inelastic \citep{Davis2016, Kilmer2014, Grossman2005, Nisbet1972}, ranging from -0.69 to -0.26. Researchers conducted the same exercise in other countries, such as South Africa, Thailand and Australia, and found that demand in those countries is also in an inelastic range \citep{Riley2020, Sukharomana2017, Van2007}. \cite{Gallet2014} observed that marijuana demand shows less responsiveness to prices than other drugs.

Our results suggests that selection into access is relevant for the intensive margin conditional in the extensive margin, and that modeling simultaneously the three stages is important. We found that marijuana is an inelastic good (the average elasticity is -0.45), there is not statistically significance regarding price heterogeneity among age groups, and the risk perception about its use is also not relevant on the intensive margin, but it is very relevant in the access and extensive margin. Moreover, we did not find heterogeneity regarding age splines in the intensive margin, but there is also relevant heterogeneity regarding the access and extensive margin. In general, we found that demographic and socioeconomic features are important to explain the three stages of marijuana demand in Colombia, all parameters estimates give intuitive results, for instance, women have less probability of having access and using marijuana, and their consumption is 45.5\% lower than consumption of men.

Regarding the counterfactual exercises, we found that legalization of marijuana implies an overall increase of the probability of use in 0.7 percentage points (p.p.), from 2.3\% pre-legalization to 3.0\% under legalization. The population group that faces the higher increase in the probability of consumption is younger individuals (20s age spline) with a low or medium risk perception about using marijuana, this is 3.8 p.p., from 13.6\% to 17.4\%. In addition, tax revenues from taxation to marijuana may fluctuate between USD 11.0 million to USD 54.2 million, this depends on the tax setting. A potential benchmark is USD 32 million, where marijuana tax is US\cent 37.8, which implies a price equal to US\cent 39.1, half the price of actual marijuana for individuals with access. 

After this introduction, we show our econometric framework in Section \ref{sec:Model}. Section \ref{sec:App} shows the results of the demand of marijuana in Colombia using information from the National Survey on the Consumption of Psychoactive Substances in 2019. Section \ref{Checks} shows results of tests regarding exclusionary restrictions and some robustness checks. Section \ref{policy} shows results of potential implications of marijuana legalization on the extensive and intensive margins, and potential revenues that the government would have from this public policy. Concluding remarks are shown in Section \ref{sec:Con}.

\section{Econometric approach}\label{sec:Model}

\subsection{Drug access}
We set $A_{im}$ indicating if individual $i$ in market $m$ has access to marijuana,
\begin{align}\label{eq:1}
    A_{im}=\begin{Bmatrix}
    1, & U_{im}^a>0\\
    0, & U_{im}^a\leq 0\\ 
    \end{Bmatrix},
\end{align}

where $U_{im}^a=\bm{w}_i^{\top}\boldsymbol{\alpha}_a+o_i\tau_a+\boldsymbol{\omega}_a+d_{im}\beta_a+V_{im}$ is the access latent variable, $i=1,\dots,N$, $m=1,\dots,M$.

Equation \ref{eq:1} defines if individual $i$ has access to marijuana in market $m$ ($A_{im} = 1$), or not ($A_{im} = 0$), if she/he has a net positive utility from it. The latter is a function of socioeconomic and demographic characteristics ($\bm{w}_i$) such as age splines (teenager or 20’s, 30’s, 40’s, and 50’s or older), socioeconomic strata (low, medium and high), years of education, gender, mental and physical health status (good or bad), a dummy variable indicating if friends or family members use marijuana, and risk perception about using marijuana (low, medium and high). The latter variable is associated with mental, physical or/and social risks. There is no legal risk for marijuana users in Colombia as it is legally allowed the personal dose of regular cannabis (up to 20 grams). The risk perception is potentially influenced by the public policy, as marketing campaigns may affect risk perception about marijuana use. We also control for a dummy variable indicating previous consumption of alcohol and cigarette ($o_i$), characteristics of the market such regional-fixed effects ($\boldsymbol{\omega}_a$), and presence of drug dealers in the neighborhood ($d_{im}$). The latter is a supply side variable that should affect access to marijuana. The location parameters are $\boldsymbol{\alpha}_a$, $\boldsymbol{\omega}_a$ and $\beta_a$, and we assume $V_{im}\sim N(0,1)$, where the variance is set to 1 due to scale identification issues.

\subsection{Drug extensive margin}
The latent variable $U^c_{im} = \bm{w}^{\top}_i \boldsymbol{\alpha}_c + \omega_c + o_i\tau_c + E_{im}$ defines drug use (extensive margin),

\begin{align}\label{eq:2}
    C_{im}=\begin{Bmatrix}
    1, & U_{im}^c>0|A_{im}=1\\
    0, & U_{im}^c\leq 0|A_{im}=1\\
    -, & A_{im}=0
    \end{Bmatrix}.
\end{align}

We observe if individual $i$ uses marijuana in market $m$ ($C_{im} = 1$), or not ($C_{im} = 0$). Individuals use marijuana if they have a net positive utility from it, conditional on having access ($A_{im} = 1$). Otherwise, individuals do not use marijuana if they do not have a net positive utility from it conditional on having access. Observe that we have missing values
regarding consumption when individuals do not have access to marijuana. This is due to individuals without access may or may not have net positive utility from marijuana use, therefore, we do not have information about their potential preferences, and consequently, this set of individuals do not contribute to identify these parameters.

The net indirect utility defining drug use depends on demographic and socioeconomic characteristics ($\bm{w}_i$), regional-fixed effects ($\omega_c$), and a dummy indicating previous consumption of alcohol and cigarette ($o_i$). The latter variable is due to the \textit{gateway hypothesis} which indicates that previous consumption of these legal substances precedes use of illicit drugs \citep{Kandel1975,Kandel1992}. Observe that we do not use price of marijuana in the extensive margin equation as due to its low price in Colombia (US\cent 83 per joint), this variable should not affect the extensive margin. We assume that $E_{im} \sim N(0, 1)$ due to scale identification issues, and the location parameters are $\boldsymbol{\alpha}_c$, $\omega_c$ and $\tau_c$.

\subsection{Drug intensive margin}

We model the consumption quantity (intensive margin), as a function of socioeconomic and demographic characteristics ($\bm{w}_i$), regional-fixed effects ($\omega_y$), price of marijuana ($p_i$), and the interaction between price and age brackets ($w_i^{\text{Age(j)}}\times p_{i}$, where Age(j) refers to j-th age bracket). The latter due to potential heterogeneity regarding price sensitivity among age splines. Observe that we have prices at individual level weighted by quality (see Appendix \ref{sec:appendix1} sections \ref{quality} and \ref{prices}), this implies that endogeneity would not be a concern due to there is no demand-supply simultaneity neither endogeneous variation in the stochastic error due to quality of marijuana \citep{jacobi2016marijuana}.

%We consider two modeling settings regarding the intensive margin. The first model is considering incidental truncation, that is, there are missing values for consumption quantity when individuals do not have access, or having access reporting not to use marijuana. The latter due to some marijuana users reporting not consumption because of social stigma.

We consider in the intensive margin equation incidental truncation, that is, there are missing values for consumption quantity when individuals do not have access, or having access reporting not to use marijuana.

\begin{align}\label{eq:3}
    Y_{im}=\begin{Bmatrix}
    \bm{w}^{\top}_i \boldsymbol{\alpha}_y + \omega_y + p_i\gamma_y + \sum_{j=1}^J w_i^{\text{Age(j)}}\times p_{i}\gamma_{yj} + W_{im}, & C_{im}=1\\
    - & C_{im}=0 \ \text{or} \ A_{im}=0
    \end{Bmatrix}.
\end{align}

Observe that the truncation setting given in equation \ref{eq:3}, %can be considered as an upper bound%, 
as we are not taking into account zero consumption. This is because some marijuana users reporting not consumption due to social stigma. 
%In our robustness checks we estimate the model taking into account individuals who report zero consumption (see equation \ref{eq:3a} in Section \ref{robustness}), this would be a lower bound in the intensive margin equation. %
The location parameters related to consumption level are $\boldsymbol{\alpha}_y$, $\omega_y$, $\gamma_y$ and $\gamma_{yj}$, and we assume $W_{im}\sim N(0, \sigma^2_y)$.

\subsection{Correlation on unobservable variables}

We model simultaneously the three stages (access, extensive and intensive margins) due to there should be unobservable variables that drive these stages. Thus, we assume $\Xi_{im} = \left[V_{im} \ E_{im} \ W_{im}\right]^{\top}\sim N_3(\bm{0}, \boldsymbol{\Sigma})$, where

\begin{align}\label{eq:4}
    \boldsymbol{\Sigma}=\begin{bmatrix}
    1 & \sigma_{ac} & \sigma_{ay}\\
    \sigma_{ca} & 1 & \sigma_{cy}\\
    \sigma_{ya} & \sigma_{yc} & \sigma_{y}^2      
    \end{bmatrix}.
\end{align}

Observe that if $\boldsymbol{\Sigma}$  is a diagonal matrix, the three stages are independent, and we can perform inference estimating each equation separately. This maybe a situation where there is no strategic search of drug dealers by users, and that more intense consumers do not make a greater effort to get drug dealers when arriving to a new market. On the other hand, if there is endogenous access, which means $\sigma_{ca}\neq 0$ and/or $\sigma_{ya}\neq 0$, we should model simultaneously these equations to get good sampling properties of our estimators. In addition, $\sigma_{yc}$ takes into account potential unobserved dependence between extensive and intensive margins conditional on selection into access.

\subsection{Estimation strategy}\label{subsec:est}

Observe that modeling the joint distribution of access, extensive and intensive margin implies to integrate over a multivariate space to recover the likelihood function. In addition, this is not a standard likelihood as incidental truncation implies that different sets of individuals contribute to different sets of parameters. Particularly, there are three different groups of individuals when taking the model setting given by equations \ref{eq:1}, \ref{eq:2} and \ref{eq:3}: all individuals ($G_1$) contribute to estimate the location parameters in the access equation, individuals who report having access ($G_2$) contribute to estimate the location parameters in the extensive margin equation and $\sigma_{ac}$, and individuals who report to use marijuana ($G_3$) contribute to estimate the parameters of the intensive margin equation, and $\sigma_{ay}$, $\sigma_{cy}$ and $\sigma_{y}^2$. 

Thus, we use data augmenting \citep{Tanner1987} to facilitate inference, we treat latent variables as parameters, such that the augmented model is,

\begin{align}\label{eq5a}
	\underbrace{\begin{bmatrix}
			{U}_{im}^a \\
			{U}_{im}^c \\
			{Y}_{im}
	\end{bmatrix}}_{\bm{T}_{im}} & = \underbrace{\begin{bmatrix}
			\bm{x}_{im}^{a\top} & \bm{0} & \bm{0}\\ 
			\bm{0} & \bm{x}_{im}^{c\top} & \bm{0}\\
			\bm{0} & \bm{0} & \bm{x}_{im}^{y\top} 
	\end{bmatrix}}_{\bm{X}_{im}} \underbrace{\begin{bmatrix}
			\boldsymbol{\theta^a}\\
			\boldsymbol{\theta^c}\\
			\boldsymbol{\theta^y}
	\end{bmatrix}}_{\boldsymbol{\theta}} + \underbrace{\begin{bmatrix}
			{V}_{im}\\
			{U}_{im}\\
			{E}_{im}
	\end{bmatrix}}_{\boldsymbol{\Xi}_{im}},
\end{align}

where $\bm{x}_{im}^{s\top}$ is the vector of regressors associated with individual $i$ in market $m$ in stage $s = \left\{\text{access, consumption, quantity}\right\}$, and $\boldsymbol{\theta}^s$ is the vector of location parameters in stage s.

The likelihood function is

\begin{align}\label{eq:6}	p(\tilde{\bm{T}}_{1},\dots,\tilde{\bm{T}}_{n}|\tilde{\bm{X}}_{1},\dots,\tilde{\bm{X}}_{n},\boldsymbol{\theta},\boldsymbol{\Sigma})&=\prod_{s=1}^{G}\prod_{i\in G_s} \left\{\mathds{1}(A_{im}=0)\mathds{1}(U_{im}^a\leq 0)+\mathds{1}(A_{im}=1)\mathds{1}(U_{im}^a > 0)\right.\nonumber\\
	&\left.\times\left[\mathds{1}(C_{im}=0)\mathds{1}(U_{im}^c\leq 0)+\mathds{1}(C_{im}=1)\mathds{1}(C_{im}^c> 0)\right]\right\}\nonumber\\ &\times\phi(\tilde{\bm{T}}_{i}|\tilde{\bm{X}}_{im}\tilde{\boldsymbol{\theta}}_{G_s},\tilde{\boldsymbol{\Sigma}}_{G_s}), 
\end{align}

where $G = 3$, $\phi(\cdot|\tilde{\bm{X}}_{im}\tilde{\boldsymbol{\theta}}_{G_s},\tilde{\boldsymbol{\Sigma}}_{G_s})$ is the density function of a normal distribution with mean $\tilde{\bm{X}}_{im}\tilde{\boldsymbol{\theta}}_{G_s}$ and variance $\tilde{\boldsymbol{\Sigma}}_{G_s}$, and $\tilde{M} = f(G_s,M)$, where $f(G_s,M)$ is a function that takes as inputs a state ($G_s$) and a
matrix ($M$), and returns as output the appropriate subset of rows and columns of $M$. For instance,
\begin{equation}
	\tilde{\bm{T}}_{im}\equiv f(G_2,\bm{T}_{im})=\begin{bmatrix}
		U_{im}^a \\
		U_{im}^c \\
		\end{bmatrix}, \ \tilde{\bm{X}}_{im}\equiv f(G_2,\bm{X}_{im})=\begin{bmatrix}
	\bm{x}_{im}^{a\top} & \bm{0} \\
	\bm{0} & \bm{x}_{im}^{c\top}\\ 
\end{bmatrix}, \nonumber
\end{equation}
\begin{equation}
	 \tilde{\boldsymbol{\theta}}_{G_2}\equiv f(G_2,\boldsymbol{\theta})=\begin{bmatrix}
		\boldsymbol{\theta}^a \\
		\boldsymbol{\theta}^c \\
	\end{bmatrix} \text{and} \
\tilde{\boldsymbol{\Sigma}}_{G_2}\equiv f(G_2,\boldsymbol{\Sigma})= \begin{bmatrix}
1 & \sigma_{ac} \\
\sigma_{ca} & 1 \\   
\end{bmatrix}.  \nonumber
\end{equation}

We use the Bayes’ rule to perform inference in our model. However, we implement
our inferential algorithm in the unidentified parameter space as getting draws from the posterior distribution in the identified parameter space has a high computational cost and inferior mixing properties than a more straightforward Gibbs sampler traversing over the unidentified space \citep[p.~118]{Rossi2005}. Thus, we set
\begin{align}\label{eq:7}
    \boldsymbol{\Omega}=\begin{bmatrix}
    \omega_a^2 & \omega_{ac} & \omega_{ay}\\
    \omega_{ca} & \omega_c^2 & \omega_{cy}\\
    \omega_{ya} & \omega_{yc} & \omega_{y}^2      
    \end{bmatrix}.
\end{align}

We implement our Gibbs sampling algorithm using standard conjugate independent priors to obtain standard conditional posterior distributions that facilitate computation. In particular, we assume that $\pi(\boldsymbol{\theta},\boldsymbol{\Omega})=N(\boldsymbol{\theta}|\boldsymbol{\theta}_0,\boldsymbol{\Theta}_0)\times \pi(\boldsymbol{\Omega}|\bm{R}_0,r_0)$, that is, a multivariate normal distribution for the location parameters and an inverse Wishart distribution for the unidentified covariance matrix. We use non-informative hyperparameters in all our exercises, that is, $\boldsymbol{\theta}_0=\bm{0}$, $\boldsymbol{\Theta}_0=\text{diag}\left\{1,000\right\}$, $\bm{R}_0=\bm{I}_{3}$ and $r_0=3+2$.

Reparameterizing the data-augmented likelihood function from equation \ref{eq:6} in terms of $\boldsymbol{\Omega}$, and using the previous prior density function, the Bayes’ rule implies that the posterior conditional distribution for the location parameters is $N(\boldsymbol{\theta}|\boldsymbol{\theta}_n,\boldsymbol{\Theta}_n)$ where $\boldsymbol{\Theta}_n=\left[\sum_{s=1}^{G}\sum_{i\in G_s}\bm{J}_{G_s}\tilde{\bm{X}}_{im}^{\top}\tilde{\boldsymbol{\Omega}}_{G_s}^{-1}\tilde{\bm{X}}_{im}\bm{J}_{G_s}^{\top}+\boldsymbol{\Theta}_0^{-1}\right]^{-1}$ and $\boldsymbol{\theta}_n=\boldsymbol{\Theta}_n\left[\sum_{s=1}^{G}\sum_{i\in {G_s}}\bm{J}_{G_s}\tilde{\bm{X}}_{im}^{\top}\tilde{\boldsymbol{\Omega}}_{G_s}^{-1}\tilde{\bm{T}}_{im}+\boldsymbol{\Theta}_0^{-1}\boldsymbol{\theta}_0\right]$, and
\begin{equation*}
	\bm{J}_{G_1} =
	\begin{bmatrix}
		\bm{I}_{\left\{H\right\}}\\
		\bm{0}\\
		\bm{0}\\
	\end{bmatrix}, 	\bm{J}_{G_2} =
	\begin{bmatrix}
		\bm{I}_{\left\{H\right\}} & \bm{0}\\
		\bm{0} & \bm{I}_{\left\{K\right\}}\\
		\bm{0} & \bm{0}\\
	\end{bmatrix} \ \text{and} \ \bm{J}_{G_3} =\begin{bmatrix}
		\bm{I}_{\left\{H\right\}} & \bm{0} & \bm{0}\\
		\bm{0} & \bm{I}_{\left\{K\right\}} & \bm{0}\\
		\bm{0} & \bm{0} & \bm{I}_{\left\{L\right\}}\\
	\end{bmatrix}, 
\end{equation*} 

where $\bm{I}_{\left\{d\right\}}$ is a $d\times d$ identity matrix, $H$, $K$ and $L$ are the dimensions of $\boldsymbol{\theta}_s$.

We follow a sequential approach \citep{chib2009estimation,li2011estimation} to get standard conditional posterior distributions that allow to recover $\boldsymbol{\Omega}$. In particular, all observations contribute to estimate $\omega_a^2$, and given the prior distribution of $\boldsymbol{\Omega}$, which implies that the prior distribution of $\omega_a^2$ is inverse gamma with parameters $r_{11,0}$ and $r_0-2$, where $r_{11,0}$ is the element 1,1-th of $\bm{R}_0$ \citep[p.~ 190]{greenberg2012introduction}, then the posterior conditional distribution of $\omega_a^2$ is $IG(r_{11,n}, r_0-2+n)$ where $r_{11,n}=\sum_{i=1}^n(T_{1,im}-X_{1,im}\boldsymbol{\theta})^2 + r_{11,0}$, $T_{1,im}$ is the first element of $\bm{T}_{im} (U^a_{im})$ and $X_{1,im}$ is the first row of $\bm{X}_{im}$.

In the next stage, we set
\begin{align*}
    \boldsymbol{\Omega}_{22}=\begin{bmatrix}
        \omega_a^2 & \omega_{ac}\\
        \omega_{ca} & \omega_c^2
    \end{bmatrix},
\end{align*}
and
\begin{equation}\label{eq:8}
    \omega_{c.1}^2=\omega_c^2-\frac{\omega_{ac}^2}{\omega_a^2}.
\end{equation}
Given the prior distribution of $\boldsymbol{\Omega}$, and a consistent partition of $\bm{R}_0$,
\begin{align*}
    \bm{R}_{22,0}=\begin{bmatrix}
        r_{11,0}^2 & r_{12,0}\\
        r_{21,0} & r_{22,0}^2
    \end{bmatrix},
\end{align*}
the prior distribution of $\omega_{c.1}^2$ is inverse gamma with parameters $r_{22.1,0}^2$ and $r_0$, where $r_{22.1,0}^2=r_{22,0}^2-r_{12,0}^2/r_{11,0}^2$.

In addition, we set 
\begin{equation}\label{eq:9}
    \omega_{ca.1}=\frac{\omega_{ca}}{\omega_a^2},
\end{equation}
and given the prior distribution of $\boldsymbol{\Omega}$, the prior distribution of $\omega_{ca.1}|\omega_{c.1}^2$ is normal with mean $r_{21,0}/r_{11,0}^2$ and variance $\omega_{c.1}^2/r_{11,0}^2$.

We calculate
\begin{equation}
    r_{22,n}=\sum_{i\in G_2}\left\{(\bm{T}_{1:2,im}-\bm{X}_{1:2,im}\boldsymbol{\theta})(\bm{T}_{1:2,im}-\bm{X}_{1:2,im}\boldsymbol{\theta})^{\top}\right\}+\bm{R}_{22,0}=\begin{bmatrix}
        r_{11,n}^2 & r_{12,n}\\
        r_{21,n} & r_{22,n}^2
    \end{bmatrix},\nonumber
\end{equation}
where $\bm{T}_{1:2,im}$ and $\bm{X}_{1:2,im}$ are the first and second rows of $\bm{T}_{im}$ and $\bm{X}_{im}$, respectively.

The posterior distribution of $\omega_{c.1}^2$ is inverse gamma with parameters $r_{22.1,n}^2$ and $r_0+|G_2|$, where $r_{22.1,n}^2=r_{22,n}^2-r_{12,n}^2/r_{11,n}^2$, and $|G_2|$ is the number of individuals in group two.

The posterior distribution of $\omega_{ca.1}$ conditional on $\omega_{c.1}^2$ is normal with mean $r_{21,n}/r_{11,n}^2$ and variance $\omega_{c.1}^2/r_{11,n}^2$.

We can recover $\boldsymbol{\Omega}_{22}$ using equation \ref{eq:9}, such that $\omega_{ca}=\omega_{ca.1}\omega_{a}^2$, and equation \ref{eq:8}, where we have that $\omega_c^2=\omega_{c.1}^2+\omega_{ca}^2/\omega_a^2$.

We set
\begin{align*}
    \boldsymbol{\Omega}=\begin{bmatrix}
        \boldsymbol{\Omega}_{22} & \boldsymbol{\Omega}_{23}\\
        \boldsymbol{\Omega}_{32}& \omega_y^2
    \end{bmatrix},
\end{align*}
where $\boldsymbol{\Omega}_{32}=\left[\omega_{ya} \ \omega_{yc}\right]$, and 
\begin{equation}\label{eq:10}
    \omega_{y.1}^2=\omega_y^2-\boldsymbol{\Omega}_{32}\boldsymbol{\Omega}_{22}^{-1}\boldsymbol{\Omega}_{23},
\end{equation}
and given the prior distribution of $\boldsymbol{\Omega}$, and a consistent partition of $\bm{R}_0$,
\begin{align*}
    \bm{R}_0=\begin{bmatrix}
        \bm{R_}{22,0} & \bm{R}_{23,0}\\
        \bm{R}_{32,0}& r_{33,0}^2
    \end{bmatrix},
\end{align*}
the prior distribution of $\omega_{y.1}^2$ is inverse gamma with parameters $r_{33.1,0}^2$ and $r_0$, where $r_{33.1,0}^2=r_{33,0}^2-\bm{R}_{32,0}\bm{R}_{22,0}^{-1}\bm{R}_{23,0}$.

Given 
\begin{equation}\label{eq:11}   \boldsymbol{\Omega}_{32.1}=\boldsymbol{\Omega}_{32}\boldsymbol{\Omega}^{-1}_{22},
\end{equation}
where the prior distribution of $\boldsymbol{\Omega}_{32.1}|\omega_{y.1}^2$ is matrix normal with mean $\bm{R}_{32,0}\bm{R}_{22,0}^{-1}$ and scale matrices $\bm{R}^{-1}_{22,0}$ and $\omega^{2}_{y.1}$.

Given
\begin{equation}
    \bm{R}_{n}=\sum_{i\in G_3}\left\{(\bm{T}_{im}-\bm{X}_{im}\boldsymbol{\theta})(\bm{T}_{im}-\bm{X}_{im}\boldsymbol{\theta})^{\top}\right\}+\bm{R}_{0}=\begin{bmatrix}
        \bm{R}_{22,n}^2 & \bm{R}_{23,n}\\
        \bm{R}_{32,n} & r_{33,n}^2
    \end{bmatrix},\nonumber
\end{equation}
The posterior distribution of $\boldsymbol{\Omega}_{32.1}$ conditional on $\omega^2_{y.1}$ is matrix normal with mean $\bm{R}_{32,n}\bm{R}^{-1}_{22,n}$ and scale matrices $\bm{R}^{-1}_{22,n}$ and $\omega^2_{y.1}$. We can recover $\boldsymbol{\Omega}$ using equation \ref{eq:11}, such that $\boldsymbol{\Omega}_{32}=\boldsymbol{\Omega}_{32.1}\boldsymbol{\Omega}_{22}$, and equation \ref{eq:10}, where we have that $\omega^2_y=\omega^2_{y.1}+\boldsymbol{\Omega}_{32}\boldsymbol{\Omega}^{-1}\boldsymbol{\Omega}_{23}$.

The posterior conditional distributions of $U_{im}^a$ and $U_{im}^c$ are truncated normal in the interval $(-\infty,0]$ if $U_{im}^l=0$, and $(0,\infty)$ if $U_{ij}^l=1$, respectively, $l=\left\{a,c\right\}$. Their conditional means are $m_{l,im}=\tilde{\bm{X}}_{l,im}\tilde{\boldsymbol{\theta}}+\tilde{\boldsymbol{\Omega}}_{l,-l}\tilde{\boldsymbol{\Omega}}_{-l,-l}^{-1}(\tilde{\bm{T}}_{-l,im}-\tilde{\bm{X}}_{-l,im}\tilde{\boldsymbol{\theta}})$, and conditional variances $\tau_{l}^2=\tilde{\omega}_{ll}^2-\tilde{\boldsymbol{\Omega}}_{l,-l}\tilde{\boldsymbol{\Omega}}_{-l,-l}^{-1}\tilde{\boldsymbol{\Omega}}_{-l,l}$, where $\tilde{\bm{T}}_{-l,im}$ is the vector $\tilde{\bm{T}}_{im}$ excluding the $l$-th component, $\tilde{\bm{X}}_{l,im}$ is the $l$-th row of matrix $\tilde{\bm{X}}_{im}$, $\tilde{\bm{X}}_{-l,im}$ is the matrix $\tilde{\bm{X}}_{im}$ without the $l$-th row, $\tilde{\boldsymbol{\Omega}}_{l,-l}$ is the $l$-th row of $\tilde{\boldsymbol{\Omega}}$ excluding the $l$-th element, $\tilde{\boldsymbol{\Omega}}_{-l,-l}$ is equal to $\tilde{\boldsymbol{\Omega}}$ excluding the $l$-th row and $l$-th column, and $\tilde{\omega}^2_{ll}$ is the $ll$ element of $\tilde{\boldsymbol{\Omega}}$.

Our econometric framework differs from other literature regarding modeling marijuana consumption in a few fundamental ways. In particular, we follow a simultaneous three-part modeling approach that takes into account that access is a necessary condition for use, such that the latter is endogenously determined with extensive and intensive margins. We also incorporate that use is a necessary condition for the intensive margin, and allow for unobserved correlation between these stages. In addition, we consider the incidental truncation issue due to missing reports when individuals report no to have access to marijuana, or when reporting access, they report not to use it. Omitting access restrictions, potential correlation on unobservable variables and/or incidental truncation may generate inconsistent estimators. There is also a clear link between our modeling strategy and the policy that we want to analyze, as marijuana legalization implies basically free access for all potential users, such that the ``breaking the law" hindrance will disappear, this is formally $P(A_{imM}=1)=1$ in our econometric framework, and will be the basis for our counterfactual exercises.

\section{Marijuana demand in Colombia}\label{sec:App}

\subsection{Data}
We apply our approach to the problem of estimating the demand for marijuana in Colombia. We leveraged individual-level data on the consumption of psychoactive substances representative of the entire country in 2019. In particular, our data come from the National Survey on the Consumption of Psychoactive Substances (\textit{ENCSPA} 2019), which is a national representative survey, performed by the government statistical department of Colombia (DANE) aiming to measure both legal and illegal substance abuse within the population. The survey randomly sampled households from several municipalities in Colombia. It targeted individuals between 12 and 65 years of age, who were selected randomly from all the household members that met the age criterion. The enumerators privately performed the survey. If the chosen person was absent during the survey, the enumerator should return later but was not allowed to change the individual to be interviewed.

We are especially interested in variables describing marijuana consumption patterns based on the literature review and data availability. Our measure of marijuana consumption takes into account quality based on the \textit{tetrahydrocannabinol} (THC) content (see Appendix \ref{sec:appendix1}, subsection \ref{quality}). In addition, we use a nearest neighbor algorithm to impute prices for those individuals who report not to consume marijuana, and consequently, who do not report average price (see Appendix \ref{sec:appendix1}, subsection \ref{prices}),\footnote{Observe that posterior estimates of our model composed by equation \ref{eq:1}, \ref{eq:2} and \ref{eq:3} does not require these prices.} and a basic counting algorithm to construct risk perception about drug use (see Appendix \ref{sec:appendix1}, subsection \ref{risk}).

Table \ref{tab:DescStats} presents the summary statistics for the key variables in our sample. We consider three different outcome variables, one for each part of the three-part model: First, whether the individual has access or not to marijuana by reporting that it would be easy for her/him to get it. Second, whether the individual is a consumer, specifically if they had consumed marijuana during the last 12 months, and finally, the quantity of marijuana consumed on average per month. The quantity consumed is measured in the number of cigarettes or joints of marijuana; although it is not specified the number of grams, there is some consistency in the sizes of common joints, which are usually about 1 gram.

%%% Descriptive statistics table 3
\begin{table}[h!]\centering \caption{Summary Statistics.\label{tab:DescStats}}
	\begin{threeparttable}
		\resizebox{0.73\textwidth}{!}{\begin{minipage}{\textwidth}
\begin{tabular}{llccccccccc}
\hline
 & & All & \multicolumn{1}{l}{} & \multicolumn{3}{c}{Access to marijuana} & \multicolumn{1}{l}{} & \multicolumn{3}{c}{Marijuana consumer} \\ \cline{3-3} \cline{5-7} \cline{9-11} 
 & & \multicolumn{1}{l}{} & \multicolumn{1}{l}{} & No & & Yes & & No & & Yes \\ \cline{5-5} \cline{7-7} \cline{9-9} \cline{11-11} 
Variable & & (1) & & (2) & & (3) & & (4) & & (5) \\ \cline{3-11} 
 & & & & & & & & & & \\
\multirow{2}{*}{Access to marijuana} & & 0.58 & & 0.00 & & 1.00 & & 0.57 & & 1.00 \\
 & & (0.49) & & (0.00) & & (0.00) & & (0.5) & & (0.00) \\
\multirow{2}{*}{Marijuana consumer} & & 0.02 & & 0.00 & & 0.04 & & 0.00 & & 1.00 \\
 & & (0.15) & & (0.00) & & (0.20) & & (0.00) & & (0.00) \\
\multirow{2}{*}{Quantity consumed} & & 1.01 & & 0.00 & & 1.75 & & 0.00 & & 43.08 \\
 & & (13.61) & & (0.00) & & (17.89) & & (0.00) & & (78.07) \\
\multirow{2}{*}{Drug dealer in neighborhood} & & 0.38 & & 0.26 & & 0.47 & & 0.38 & & 0.50 \\
 & & (0.49) & & (0.44) & & (0.50) & & (0.48) & & (0.50) \\
\multirow{2}{*}{Alcohol and tobacco user} & & 0.32 & & 0.22 & & 0.40 & & 0.31 & & 0.86 \\
 & & (0.47) & & (0.41) & & (0.49) & & (0.46) & & (0.35) \\
\multirow{2}{*}{Price of marijuana} & & 0.83 & & 0.86 & & 0.80 & & 0.83 & & 0.84 \\
 & & (0.44) & & (0.48) & & (0.40) & & (0.43) & & (0.54) \\
\multirow{2}{*}{Female} & & 0.58 & & 0.67 & & 0.52 & & 0.59 & & 0.25 \\
 & & (0.49) & & (0.47) & & (0.50) & & (0.49) & & (0.44) \\
\multirow{2}{*}{Years of education} & & 11.8 & & 11.48 & & 12.04 & & 11.79 & & 12.34 \\
 & & (4.24) & & (4.46) & & (4.04) & & (4.25) & & (3.75) \\
\multirow{2}{*}{Worker} & & 0.58 & & 0.53 & & 0.61 & & 0.58 & & 0.59 \\
 & & (0.49) & & (0.5) & & (0.49) & & (0.49) & & (0.49) \\
\multirow{2}{*}{Good mental health} & & 0.79 & & 0.81 & & 0.77 & & 0.79 & & 0.70 \\
 & & (0.41) & & (0.39) & & (0.42) & & (0.41) & & (0.46) \\
\multirow{2}{*}{Good physical health} & & 0.77 & & 0.75 & & 0.77 & & 0.76 & & 0.79 \\
 & & (0.42) & & (0.43) & & (0.42) & & (0.42) & & (0.41) \\
Marijuana users in network & & 0.36 & & 0.19 & & 0.49 & & 0.35 & & 0.94 \\
 & & (0.48) & & (0.39) & & (0.50) & & (0.48) & & (0.24) \\
Age 20's or younger & & 0.34 & & 0.30 & & 0.38 & & 0.34 & & 0.67 \\
 & & (0.47) & & (0.46) & & (0.48) & & (0.47) & & (0.47) \\
Age 30's & & 0.21 & & 0.19 & & 0.23 & & 0.21 & & 0.20 \\
 & & (0.41) & & (0.4) & & (0.42) & & (0.41) & & (0.40) \\
Age 40's & & 0.17 & & 0.17 & & 0.16 & & 0.17 & & 0.08 \\
 & & (0.37) & & (0.38) & & (0.37) & & (0.38) & & (0.27) \\
Age 50's or older & & 0.28 & & 0.34 & & 0.23 & & 0.28 & & 0.05 \\
 & & (0.45) & & (0.47) & & (0.42) & & (0.45) & & (0.23) \\
Risk perception of usage: & & \multicolumn{1}{l}{} & \multicolumn{1}{l}{} & \multicolumn{1}{l}{} & \multicolumn{1}{l}{} & \multicolumn{1}{l}{} & \multicolumn{1}{l}{} & \multicolumn{1}{l}{} & \multicolumn{1}{l}{} & \multicolumn{1}{l}{} \\
Low & & 0.04 & & 0.05 & & 0.03 & & 0.03 & & 0.09 \\
 & & (0.19) & & (0.21) & & (0.16) & & (0.18) & & (0.29) \\
Medium & & 0.05 & & 0.03 & & 0.07 & & 0.05 & & 0.30 \\
 & & (0.22) & & (0.17) & & (0.25) & & (0.21) & & (0.46) \\
High & & 0.91 & & 0.92 & & 0.90 & & 0.92 & & 0.61 \\
 & & (0.28) & & (0.26) & & (0.30) & & (0.27) & & (0.49) \\
Socio-economic Strata: & & & & & & & & & & \\
Low & & 0.64 & & 0.65 & & 0.64 & & 0.65 & & 0.56 \\
 & & (0.48) & & (0.48) & & (0.48) & & (0.48) & & (0.5) \\
Medium & & 0.32 & & 0.31 & & 0.32 & & 0.32 & & 0.38 \\
 & & (0.47) & & (0.46) & & (0.47) & & (0.47) & & (0.49) \\
High & & 0.08 & & 0.08 & & 0.08 & & 0.08 & & 0.13 \\
 & & (0.28) & & (0.28) & & (0.28) & & (0.28) & & (0.33) \\
 & & & & & & & & & & \\
Sample size & & 49,414 & & 20,909 & & 28,505 & & 48,255 & & 1,159 \\ \hline
\end{tabular}
				\begin{tablenotes}[para,flushleft]
	\footnotesize \textit{Notes}: Standard deviations in parenthesis. This table presents descriptive statistics for ENCSPA 2019 regarding access and consumption of Marijuana in Colombia, as well as control variables. Column 1 shows the information for the entire sample. Columns 2 and 3 show the information for individuals without and with access to marijuana. Columns 4 and 5 show the information for not consumers and consumers, respectively. Individuals who are consumers should have access; this is internally consistent for more than 98\% of the population and inputted for the remaining 2\%. Prices are shown in 2019 USD and were converted using the average exchange rate in 2019 (3,274 USD/COP).\\	
	\textit{Source}: Authors' construction using ENCSPA data.
	\end{tablenotes}
	\end{minipage}}
\end{threeparttable}
\end{table}

Our sample population consists of 49,414 individuals. Less than 60\% report having access (row 1, column 1); of those reporting having access, approximately 4\% report consuming marijuana during the last 12 months (row 2, column 3). Finally, the consumers take 23 joints of marijuana per month on average, with an average price of USD 0.84 (rows 3 and 6, column 5). By definition, and aiming to achieve consistency in the model, individuals who are consumers should have access; this is internally consistent in the data for more than 98\% of the population and manually inputted for the remaining 2\%\footnote{We replace access equal to one if the person is a consumer for two main reasons. First, it is not intuitive for people who are current users to report that they cannot get marijuana so it is likely an error in responding on the part of these individuals or misreporting due to stigma. Second, for the model to be internally consistent by construction, all users must have access. To review possible changes to the definition of the access measure, we conducted multiple exercises to validate the robustness of the results.} The representative individual in the survey is a female with complete secondary education in her 20s living in a low socioeconomic stratum. More than half of the people are workers, almost 80\% report having good mental health and 77\% report having 
good physical health, 91\% have a high-risk perception of the use of marijuana, and 36\% report having a peer (family or friend) who is a consumer of marijuana. Approximately, 38\% of the sample declares to have a drug dealer in the neighborhood, and 32\% report to have consumed alcohol and cigarette.

\subsection{Model results}

We perform inference of the model in system \ref{eq5a} running 6,000 iterations with a burn-in equal to 1,000 and a thin parameter equal to 5, thus we have 1,000 effective posterior draws. We compute several diagnostics to assess the convergence and stationarity of the posterior chains. In general, the posterior chains look good. Particularly, all location parameters have dependence factors that are less than 5, actually most of them less than 2, using \cite{Raftery1992}'s diagnostic, with a 95\% probability of obtaining an estimate in the interval $2.5\% \pm 1.0\%$. Regarding \cite{Heidelberger1983}'s and \cite{Geweke1992}'s tests at 5\% significance level, all parameter estimates pass the former test, and 147 out of 163 location parameters passed the latter test. The former uses the Cramer-von-Mises statistic to test the null hypothesis that the sampled values come from a stationary distribution, and the latter tests for equality of the posterior means using the first 10\% and the last 50\% of the Markov chains. Regarding the scale parameters, we have that all parameters have a dependence factor less than 5, and one out of 4 does not passed the conventional values of the \cite{Heidelberger1983} and \cite{Geweke1992} diagnostics.

%%%%%%% Posterior results of location parameters
\begin{table}[h!]\centering \caption{Posterior results of location parameters: Marijuana demand in Colombia.\label{tab:6}}
	\begin{threeparttable}
		\resizebox{0.95\textwidth}{!}{\begin{minipage}{\textwidth}
\begin{tabular}{l c c c c c c c}\hline
\multirow{4}{*}{Variable}  & \multicolumn{3}{c}{Univariate} &  & \multicolumn{3}{c}{Multivariate} \\
\cline{2-4} \cline{6-8}
 & Access & \multicolumn{2}{c}{Margin} & & Access & \multicolumn{2}{c}{Margin} \\
 \cline{3-4} \cline{7-8}
 & & Extensive & Intensive &  &  & Extensive & Intensive \\
 \cline{2-4} \cline{6-8}
& (1) & (2) & (3) & & (4) & (5) & (6) \\
\hline
Exclusionary restrictions	&		&		&		&		&		&		&		\\
\multirow{2}{*}{\quad  Drug dealer in neighborhood}	&	\textbf{0.438}	&		&		&		&	\textbf{0.433}	&		&		\\
	&	(0.014)	&		&		&		&	(0.014)	&		&		\\
\multirow{2}{*}{\quad  Alcohol and cigarette use}	&	\textbf{0.374}	&	\textbf{0.949}	&		&		&	\textbf{0.369}	&	\textbf{0.931}	&		\\
	&	(0.014)	&	(0.041)	&		&		&	(0.014)	&	(0.045)	&		\\
Age	&		&		&		&		&		&		&		\\
\multirow{2}{*}{\quad 30s}	&	\textbf{-0.095}	&	\textbf{-0.389}	&	0.241	&		&	\textbf{-0.094}	&	\textbf{-0.378}	&	0.457	\\
	&	(0.018)	&	(0.044)	&	(1.294)	&		&	(0.018)	&	(0.044)	&	(1.302)	\\
\multirow{2}{*}{\quad 40s}	&	\textbf{-0.217}	&	\textbf{-0.611}	&	0.608	&		&	\textbf{-0.216}	&	\textbf{-0.617}	&	0.534	\\
	&	(0.019)	&	(0.061)	&	(1.982)	&		&	(0.019)	&	(0.054)	&	(1.899)	\\
\multirow{2}{*}{\quad 50s and older}	&	\textbf{-0.404}	&	\textbf{-1.050}	&	-3.743	&		&	\textbf{-0.403}	&	\textbf{-1.019}	&	-2.995	\\
	&	(0.017)	&	(0.064)	&	(2.421)	&		&	(0.017)	&	(0.066)	&	(2.197)	\\
Strata	&		&		&		&		&		&		&		\\
\multirow{2}{*}{\quad Medium}	&	\textbf{-0.048}	&	\textbf{0.086}	&	-0.045	&		&	\textbf{-0.049}	&	\textbf{0.090}	&	-0.067	\\
	&	(0.015)	&	(0.039)	&	(0.098)	&		&	(0.014)	&	(0.039)	&	(0.099)	\\
\multirow{2}{*}{\quad High}	&	\textbf{-0.049}	&	\textbf{0.134}	&	-0.198	&		&	\textbf{-0.051}	&	\textbf{0.130}	&	-0.199	\\
	&	(0.024)	&	(0.056)	&	(0.140)	&		&	(0.023)	&	(0.063)	&	(0.149)	\\
Risk perception marijuana use	&		&		&		&		&		&		&		\\
\multirow{2}{*}{\quad Medium}	&	\textbf{0.536}	&	-0.112	&	\textbf{-0.573}	&		&	\textbf{0.520}	&	-0.112	&	-0.204	\\
	&	(0.043)	&	(0.078)	&	(0.162)	&		&	(0.044)	&	(0.081)	&	(0.300)	\\
\multirow{2}{*}{\quad High}	&	\textbf{0.288}	&	\textbf{-0.793}	&	\textbf{-0.644}	&		&	\textbf{0.279}	&	\textbf{-0.790}	&	-0.327	\\
	&	(0.032)	&	(0.070)	&	(0.152)	&		&	(0.034)	&	(0.072)	&	(0.247)	\\
\multirow{2}{*}{Years of education}	&	\textbf{0.017}	&	\textbf{-0.012}	&	\textbf{-0.084}	&		&	\textbf{0.017}	&	\textbf{-0.012}	&	\textbf{-0.069}	\\
	&	(0.002)	&	(0.005)	&	(0.014)	&		&	(0.002)	&	(0.005)	&	(0.015)	\\
\multirow{2}{*}{Female}	&	\textbf{-0.308}	&	\textbf{-0.459}	&	\textbf{-0.524}	&		&	\textbf{-0.310}	&	\textbf{-0.449}	&	\textbf{-0.607}	\\
	&	(0.013)	&	(0.040)	&	(0.104)	&		&	(0.013)	&	(0.037)	&	(0.146)	\\
\multirow{2}{*}{Good mental health}	&	\textbf{-0.127}	&	\textbf{-0.193}	&	0.086	&		&	\textbf{-0.129}	&	\textbf{-0.192}	&	0.052	\\
	&	(0.017)	&	(0.041)	&	(0.103)	&		&	(0.016)	&	(0.044)	&	(0.107)	\\
\multirow{2}{*}{Good physical health}	&	0.010	&	-0.037	&	-0.152	&		&	0.009	&	-0.033	&	-0.110	\\
	&	(0.016)	&	(0.044)	&	(0.114)	&		&	(0.016)	&	(0.051)	&	(0.105)	\\
\multirow{2}{*}{Marijuana users in network}	&	\textbf{0.694}	&	\textbf{0.981}	&	\textbf{0.384}	&		&	\textbf{0.695}	&	\textbf{0.971}	&	\textbf{0.619}	\\
	&	(0.014)	&	(0.050)	&	(0.183)	&		&	(0.013)	&	(0.052)	&	(0.306)	\\
\multirow{2}{*}{Worker}	&	\textbf{0.125}	&	\textbf{-0.098}	&	0.160	&		&	\textbf{0.127}	&	\textbf{-0.094}	&	0.151	\\
	&	(0.014)	&	(0.037)	&	(0.092)	&		&	(0.014)	&	(0.037)	&	(0.096)	\\
Price 	&		&		&		&		&		&		&		\\
\multirow{2}{*}{\quad $\log\left\{\text{price of marijuana}\right\}$}	&		&		&	\textbf{-0.496}	&		&		&		&	\textbf{-0.445}	\\
	&		&		&	(0.094)	&		&		&		&	(0.101)	\\
\multirow{2}{*}{\quad Age 30s $\times \log\left\{\text{price of marijuana}\right\}$}	&		&		&	0.002	&		&		&		&	-0.033	\\
	&		&		&	(0.168)	&		&		&		&	(0.167)	\\
\multirow{2}{*}{\quad Age 40s $\times \log\left\{\text{price of marijuana}\right\}$}	&		&		&	-0.051	&		&		&		&	-0.046	\\
	&		&		&	(0.257)	&		&		&		&	(0.245)	\\
\multirow{2}{*}{\quad Age 50s and older $\times \log\left\{\text{price of marijuana}\right\}$}	&		&		&	0.447	&		&		&		&	0.337	\\
	&		&		&	(0.305)	&		&		&		&	(0.283)	\\
Constant	&	\textbf{-0.540}	&	\textbf{-1.693}	&	\textbf{6.326}	&		&	\textbf{-0.527}	&	\textbf{-1.675}	&	\textbf{5.043}	\\
	&	(0.057)	&	(0.166)	&	(0.888)	&		&	(0.061)	&	(0.159)	&	(1.352)	\\
Regional-fixed effects	&	$\checkmark$	&	$\checkmark$	&	$\checkmark$	&		&	$\checkmark$	&	$\checkmark$	&	$\checkmark$	\\

\hline
Sample size	&	49,414	&	28,505	&	1,159	&		&	49,414	&	49,414	&	49,414\\
\hline
\end{tabular}
				\begin{tablenotes}[para,flushleft]
	\footnotesize \textit{Notes}: Bold font indicates statistically significant variables. Columns (1), (2) and (3) show posterior results of univariate models, and columns (4), (5) and (6) show posterior results of multivariate model. Columns (1) and (4) show results of the access equation, columns (2) and (5) show results of the extensive margin equations, and columns (3) and (6) show results of the intensive margin equation. There are 38 region fixed effects in each equation. There are meaningful differences in the intensive margin equation due to a statistically significant unobserved co-variation between the access and intensive margin equations (see Table \ref{tab:6a}). \\
	\textit{Source}: Authors' construction.
	\end{tablenotes}
	\end{minipage}}
\end{threeparttable}
\end{table}

Table \ref{tab:6} reports the posterior estimates. Columns labeled univariate show the posterior results of estimating univariate models, that is, probit models for the access and extensive margin equations, and a linear model for the logarithm of marijuana consumption (intensive margin). Columns labeled multivariate show the posterior results of modeling simultaneously equations \ref{eq:1}, \ref{eq:2} and \ref{eq:3} taking into account truncation.

We observe that univariate and multivariate models give similar results regarding the access and extensive margin equations. This is due to Table \ref{tab:6a} suggesting that these two equations are exogenous. However, we observe in Table \ref{tab:6} that posterior estimates of the univariate and multivariate models are different regarding the intensive margin (see columns (3) and (6)). This is explained by the fact that there is endogeneity between access and intensive margin. In particular, Table \ref{tab:6a} indicates that the unobserved co-variation between these equations is statistically significant. This suggests that conditional on the extensive margin, more frequent marijuana consumers make higher effort to have access to this drug.

Columns (1) and (4) in Table \ref{tab:6} show that having a drug dealer in the neighborhood increases the probability of having access to marijuana. In particular, the probability of having access to marijuana for the representative individual based on the descriptive statistics (see the last paragraph of the previous subsection) increases by 16.9 percentage points due to presence of drug dealer in the neighborhood, that is, from 36.5\% to 53.4\%, given the results in column (4). In addition, the probability of having access is lower for women, individuals who declare to have good mental health status, and are older. On the other hand, individuals who live in a low socioeconomic stratum, have a high or medium risk perception about marijuana use, who have more years of education, friends or family members that consume marijuana, and work, have a higher probability of having access to this drug.

We observe in columns (2) and (5) in Table \ref{tab:6} that previous consumption of alcohol and cigarette increases the probability of consumption. This is evidence for the gateway drug hypothesis. In addition, the probability of use increases with socioeconomic strata and having friends or family members who also consume. On the other hand, the probability of use decreases with age, risk perception, being female, having good mental health and being a worker. The results in Table \ref{tab:6} allows to predict the potential significant effects of a public policy that increases the risk perception about using marijuana. For instance, the posterior estimates in column (5) indicates that the probability of using marijuana is equal to 40.5\% for a man in his 20s that works, who has 12 years of education, consumes alcohol and cigarettes, lives in a low socioeconomic stratum, whose mental and physical health status is good, has friends who consume marijuana and has a low risk perception about using marijuana. On the other hand, this probability is equal to 15.2\% for the same individual, except that his risk perception about marijuana use is high, that is, there is a decrease of 25.3 percentage points due to changing the risk perception. 

Columns (3) and (6) in Table \ref{tab:6} show that the marijuana is an inelastic good. Particularly, column (6) shows that this elasticity is on average equal to -0.45, and is statistically significant; this agrees with previous literature \citep{Gallet2014,Davis2016,Sukharomana2017,Riley2020}. We also observe in this column that there is not statistically significant heterogeneity regarding price sensitivity between age groups. This is relevant from a public policy perspective as a valid concern regarding marijuana legalization is the implications of price variations on young individuals. The multivariate setting also suggests that the effect of risk perception is through the extensive margin, rather than directly on the intensive margin, as is suggested by the univariate modeling framework. This is, conditional on the extensive margin, the risk perception does not have any effect on the intensive margin. Finally, we see that one additional year of education decreases marijuana consumption by 6.9\%, and women consume 45.5\% ($\exp(-0.607)-1$) less than men, whereas having a network where some individuals consume marijuana increase marijuana consumption by 85.7\% ($\exp(0.619)-1$). All these variables are statistically significant.

%%%%%%%%%%%%% Posterior results of scale parameters:

    \begin{table}[h!]\centering \caption{Posterior results of scale parameters: Marijuana demand in Colombia.\label{tab:6a}}
	\begin{threeparttable}
		\resizebox{0.8\textwidth}{!}{\begin{minipage}{\textwidth}
				\begin{tabular}{l c c c c c}\hline
Posterior measures	&	Univariate	&	\multicolumn{4}{c}{Multivariate}							\\
\hline											
	&	$\sigma_{y}^2$	&	$\sigma_{ca}$	&	$\sigma_{ya}$	&	$\sigma_{yc}$	&	$\sigma_{y}^2$	\\
Mean	&	\textbf{2.103}	&	8.21.E-04	&	\textbf{7.182}	&	-0.103	&	\textbf{2.845}	\\
Standart deviation	&	(0.092)	&	(0.010)	&	(0.407)	&	(0.129)	&	(0.131)	\\

%95\% credible interval	&	[1.927, 2.276]	&	[-0.019, 0.021] 	&	[6.075, 8.145]	&	[-0.311, 0.117]	&	[2.606, 3.113]	\\
%\hline
%Sample size & 1,159  & 49,414 & 49,414 & 49,414 & 49,414\\
\hline
				\end{tabular}
				\begin{tablenotes}[para,flushleft]
	\footnotesize \textit{Notes}: Bold font indicates statistically significant variables. Columns labeled multivariate show posterior estimates of the identified covariance matrix. Column labeled univariate shows the posterior results of the variance of the intensive margin equation. The sample size in the univariate model is 1,159, and the sample size in the multivariate model is 49,414.\\
 The unobserved co-variation between the access and the intensive margin is statistically significant. The univariate model of the intensive margin shows a lower variance than the multivariate model.\\	
	\textit{Source}: Authors' construction.
	\end{tablenotes}
	\end{minipage}}
\end{threeparttable}
\end{table}

\begin{comment}
\begin{table}[h!]\centering \caption{Posterior results of scale parameters: Marijuana demand in Colombia.\label{tab:6a}}
	\begin{threeparttable}
		\resizebox{0.35\textwidth}{!}{\begin{minipage}{\textwidth}
				\begin{tabular}{c c c c}\hline
Parameter &	2.5\%	&	50\%	&	97.5\% \\
\hline
& \multicolumn{3}{c}{Multivariate}\\
\hline
$\sigma_{ca}$ &	-0.02	&	0.00	&	0.02 \\
$\sigma_{ya}$ &	7.28	&	7.93	&	8.60 \\
$\sigma_{yc}$ &	-0.19	&	0.00	&	0.17 \\
$\sigma_{y}^2$ &	2.85	&	3.11	&	3.38 \\
					\hline
& \multicolumn{3}{c}{Univariate}\\
\hline
$\sigma_{y}^2$ &	1.93	&	2.10	&	2.28 \\
\hline
				\end{tabular}
				\begin{tablenotes}[para,flushleft]
	\footnotesize \textit{Notes}: Columns labeled multivariate show posterior estimates of the identified covariance matrix. Columns labeled univariate shows the posterior results of the variance of the intensive margin equation.\\
 There is a 95\% probability that the unobserved co-variation between the access equation and the intensive margin is between 7.28 and 8.60. The univariate model of the intensive margin shows a lower variance than the multivariate model. \\	
	\textit{Source}: Authors' construction.
	\end{tablenotes}
	\end{minipage}}
\end{threeparttable}
\end{table}
\end{comment}

\section{Statistical checks}\label{Checks}

\subsection{Exclusionary restrictions}

It is well known that we can achieve identification of causal effects in nonlinear models without exclusion restrictions \citep{mcmanus1992common}. However, exclusion restrictions improve inference due to reducing variability of estimates because of data variability \citep{munkin2003bayesian}. We exclusively use presence of drug dealers in the neighborhood in the access equation as this variable affects drug supply which helps to identify demand parameters. Particularly, we would expect that presence of drug dealers would positively affect the probability of an individual having access to marijuana. We argue that the effect of this variable on the extensive and intensive margins should be just through the access. 

However, if this supply side variable affects directly the net utility of using marijuana, then our exclusion restriction would not be valid. We try to test this restriction using the subset of individuals who were offered marijuana, as a consequence, they do not have to search for this drug, which means that they are relatively free of the selection issue. We check the statistical significance of this supply side variable running a probit model on the access equation in this subsample. Column 1 of Table \ref{tab:exclusion} shows the posterior results using a non-informative normal prior distribution, the number of iterations of the Gibbs sampler is 6,000, a burn-in equal to 1,000, and a thin parameter equal to 5. We observe that the presence of drug dealer is not statistically significant in the extensive margin equation. Hence, this result suggests that ``drug dealer in neighborhood" has validity as an exclusionary restriction.

We do not use presence of drug dealer in the neighborhood neither if an individual has consumed alcohol and cigarettes any time in her/his life in the intensive margin equation. The \textit{gateway drug hypothesis} would support the latter variable as pattern of legal substance use would precede the use of illicit substances \citep{Kandel1975,Kandel1992}. This means that individuals who have consumed substances like alcohol and nicotine would have a higher probability of accessing and using marijuana. However, past or present consumption of these substances should not directly affect the intensive margin. Column 2 of Table \ref{tab:exclusion} shows the posterior results of estimating the intensive margin of marijuana, that is, the logarithm of quantity, as function of these two variables as well as all regressors in equations \ref{eq:3} using the subset of individuals that were offered marijuana, that is, the set of individuals that presumably is exogenous to the access equation. Given that we still have endogeneity due to the extensive margin in this set of individual, this is not a formal test of exclusionary restrictions. However, the fact that presence of drug dealer in the neighborhood, and consumption of alcohol and cigarettes are not statistical significant to explain the intensive margin, suggests that these variables have some validity as exclusionary restrictions.

Finally, we have included marijuana price in the intensive margin equation, but not in the access and extensive margin equations. Although this is not necessary for identification, we consider that excluding price from these equations make sense. First, the argument to exclude marijuana price from the access equation is that individuals who do not have access are unlikely to know the price of the marijuana they would obtain. Second, exclusion of marijuana price from the extensive margin equation is due to marijuana being very cheap in Colombia (US\cent 83), consequently, price is not a barrier to define the extensive margin.

%%%%% Exclusionary restrictions validation:
\begin{table}[h!]\centering \caption{Exclusionary restrictions validation: Marijuana demand in Colombia.\label{tab:exclusion}}
	\begin{threeparttable}
		\resizebox{0.62\textwidth}{!}{\begin{minipage}{\textwidth}
\begin{tabular}{l c c}\hline
\multirow{3}{*}{Variable}  & \multicolumn{2}{c}{Margin} \\
\cline{2-3}
 & Extensive & Intensive \\ 
& (1) & (2) \\
\hline
Exclusionary restrictions	&		&		\\
\multirow{2}{*}{\quad  Drug dealer in neighborhood}	&	0.065	&	0.180	\\
	&	(0.041)	&	(0.098)	\\
\multirow{2}{*}{\quad  Alcohol and cigarette use}	&	\textbf{0.837}	&	0.089	\\
	&	(0.045)	&	(0.133)	\\
Age	&		&		\\
\multirow{2}{*}{\quad 30s}	&	\textbf{-0.339}	&	0.020	\\
	&	(0.050)	&	(1.473)	\\
\multirow{2}{*}{\quad 40s}	&	\textbf{-0.498}	&	0.334	\\
	&	(0.063)	&	(2.122)	\\
\multirow{2}{*}{\quad 50s and older}	&	\textbf{-0.973}	&	-4.280	\\
	&	(0.074)	&	(2.400)	\\
Strata	&		&		\\
\multirow{2}{*}{\quad Medium}	&	\textbf{0.098}	&	-0.023	\\
	&	(0.044)	&	(0.105)	\\
\multirow{2}{*}{\quad High}	&	0.066	&	-0.047	\\
	&	(0.060)	&	(0.153)	\\
Risk perception marijuana use	&		&		\\
\multirow{2}{*}{\quad Medium}	&	-0.169	&	\textbf{-0.665}	\\
	&	(0.088)	&	(0.176)	\\
\multirow{2}{*}{\quad High}	&	\textbf{-0.817}	&	\textbf{-0.707}	\\
	&	(0.080)	&	(0.165)	\\
\multirow{2}{*}{Years of education}	&	\textbf{-0.014}	&	\textbf{-0.085}	\\
	&	(0.006)	&	(0.015)	\\
\multirow{2}{*}{Female}	&	\textbf{-0.422}	&	\textbf{-0.520}	\\
	&	(0.043)	&	(0.108)	\\
\multirow{2}{*}{Good mental health}	&	\textbf{-0.190}	&	0.203	\\
	&	(0.046)	&	(0.109)	\\
\multirow{2}{*}{Good physical health}	&	-0.031 &		\textbf{-0.270}	\\
	&	(0.049)	 &	(0.122) \\	
\multirow{2}{*}{Marijuana users in network}	&	\textbf{0.944}	&	0.230	\\
	&	(0.062)	&	(0.208)	\\
\multirow{2}{*}{Worker}	&	\textbf{-0.103}	&	0.180	\\
	&	(0.041)	&	(0.097)	\\
Price 	&		&		\\
\multirow{2}{*}{\quad $\log\left\{\text{price of marijuana}\right\}$}	&		&	\textbf{-0.526}	\\
	&		&	(0.100)	\\
\multirow{2}{*}{\quad Age 30s $\times \log\left\{\text{price of marijuana}\right\}$}	&		&	0.028	\\
	&		&	(0.191)	\\
\multirow{2}{*}{\quad Age 40s $\times \log\left\{\text{price of marijuana}\right\}$}	&		&	-0.012	\\
	&		&	(0.275)	\\
\multirow{2}{*}{\quad Age 50s and older $\times \log\left\{\text{price of marijuana}\right\}$}	&		&	0.526	\\
	&		&	(0.303)	\\
Constant	&	\textbf{-1.551}	&	\textbf{6.637}	\\
	&	(0.195)	&	(0.931)	\\
Regional-fixed effects	&	$\checkmark$	&	$\checkmark$	\\

\hline
Sample size	&	12,994	&	1,043	\\
\hline
\end{tabular}
				\begin{tablenotes}[para,flushleft]
	\footnotesize \textit{Notes}: Bold font indicates statistically significant variables. This table shows the posterior means and standard deviation (in parenthesis) of parameters of the extensive margin (column 1) and intensive margin (column 2) using the subset of individuals that were offered marijuana. These results suggest plausibility of the exclusionary restrictions. \\
	\textit{Source}: Authors' construction.
	\end{tablenotes}
	\end{minipage}}
\end{threeparttable}
\end{table}

\subsection{Robustness checks}\label{robustness}

We perform several exercises to check the robustness of the posterior estimates of our baseline specification. First, as the definition of access is very relevant in our analysis, we check our results with another definition of access. In particular, our baseline definition of access is equal to one if an individual responds that is easy to get marijuana, and zero in case that responds that it is difficult, impossible or does not know how to get it. This implies that 58\% of our sample has access to marijuana. In this exercise, we relax the access definition by including within the individuals who have access those who report that it would be easy or difficult to get marijuana. This new definition of access implies that 67\% of individuals in the sample have access (see Table \ref{tab:DescStatsRobust} in the Appendix). We reran our baseline model using this new definition of the accessibility, the results can be seen in the second column of tables \ref{tab:rob1}, \ref{tab:rob2} and \ref{tab:rob3}, where we show the posteriors estimates for the access, extensive and intensive equations, respectively.  In general, we obtain qualitatively similar results in the alternative scenario relaxing the access definition (column 2) compared to the baseline estimates (column 1) in the three stages. In most of the posterior estimates there are not statistically significant differences in both exercises, except that individuals in high strata do not have statistically significant differences compared with individuals in low strata regarding access in this new set of estimates (see Table \ref{tab:rob1}).

We also estimate our baseline model using the subset of uni-personal households. This is because lying is a valid concern when modeling demand for illicit drugs due to, for instance, social stigma \citep{lloyd2013stigmatization}. We guess that individuals who live alone have less incentives to lie regarding marijuana use. Column (3) in tables  \ref{tab:rob1}, \ref{tab:rob2} and \ref{tab:rob3} show the results using this sub-sample. We observe again that there are not statistically significant differences compared to the baseline exercise in the access and extensive margin, except that now there are not statistically significant differences regarding strata or being a worker. In addition, years of education is not statistically relevant in the extensive margin. However, we observe some intriguing results in the intensive margin estimates. In particular, three very robust regressors in all the estimations are not statistically significant in this subset: female, marijuana users in the network and price. Although their coefficients have the expected sign. We suspect that can be a power issue due to being just 233 marijuana consumers in this sub-sample.      

We also include interaction effects between age splines and risk perception about marijuana consumption in our main specification. We perform this to identify potential heterogeneous effects in this variable among age groups, thus thinking about marketing campaigns targeting young adults to curve marijuana consumption through risk perception. We observe in column (4) of Table \ref{tab:rob3} that these variables are not statistically significant, and in general, we get very similar results in this alternative specification compared to the baseline exercise.  

We get our price measure in the baseline estimation calculating expenditure in marijuana over quantity. The advantage of this measure is that takes implicitly quality into account when an individual buys different types of marijuana. However, the survey asks directly individuals about price, there is the question ``Do you know how much a marijuana cigarette or joint costs? We estimate our model using this alternative measure of price, which implies calculating again the quantity weighted by THC. Column (5) in tables \ref{tab:rob1}, \ref{tab:rob2} and \ref{tab:rob3} show the results. It seems that our results are robust to the price measure, the price elasticity is numerically lower under the alternative price compared to the baseline exercise, but there is not statistically significant differences. 

A potential endogeneity issue that we did not take into account in our main specification is self-perception about health status. This variable may be considered endogenous \citep{jacobi2016marijuana}, thus we estimate the baseline specification without the mental and physical self-perception of health status. The results can be seen in column (6) of tables \ref{tab:rob1}, \ref{tab:rob2} and \ref{tab:rob3}. The results are very similar comparing this exercise with the baseline specification, except that medium strata do not have statistically significant differences with the low strata in the extensive margin in this new setting.  

%%%%%%% Posterior results of location parameters: Access equation %%%%%%%%%%%%%%
\begin{table}[h!]\centering \caption{Robustness checks: Posterior results of location parameters in access of marijuana demand in Colombia.\label{tab:rob1}}
	\begin{threeparttable}
		\resizebox{1\textwidth}{!}{\begin{minipage}{\textwidth}
\begin{tabular}{l c c c c c c }\hline
\multirow{3}{*}{Variable}	&	Baseline	&	Access	&	Uni-personal	&	Interaction	&	Price	&	Perception	\\
	&	Specification	&	definition	&	household	&	Age and risk	&	definition	&	health status	\\
	&	(1)	&	(2)	&	(3)	&	(4)	&	(5)	&	(6)	\\
\hline													
Exclusionary restrictions	&		&		&		&		&		&		\\
\multirow{2}{*}{\quad  Drug dealer in neighborhood}	&	\textbf{0.433}	&	\textbf{0.366}	&	\textbf{0.441}	&	\textbf{0.433}	&	\textbf{0.432}	&	\textbf{0.433}	\\
	&	(0.014)	&	(0.014)	&	(0.043)	&	(0.014)	&	(0.014)	&	(0.013)	\\
\multirow{2}{*}{\quad  Alcohol and cigarette use}	&	\textbf{0.369}	&	\textbf{0.330}	&	\textbf{0.316}	&	\textbf{0.369}	&	\textbf{0.365}	&	\textbf{0.374}	\\
	&	(0.014)	&	(0.015)	&	(0.042)	&	(0.015)	&	(0.015)	&	(0.014)	\\
Age	&		&		&		&		&		&		\\
\multirow{2}{*}{\quad 30s}	&	\textbf{-0.094}	&	\textbf{-0.093}	&	\textbf{-0.214}	&	\textbf{-0.093}	&	\textbf{-0.092}	&	\textbf{-0.086}	\\
	&	(0.018)	&	(0.019)	&	(0.065)	&	(0.018)	&	(0.018)	&	(0.018)	\\
\multirow{2}{*}{\quad 40s}	&	\textbf{-0.216}	&	\textbf{-0.195}	&	\textbf{-0.386}	&	\textbf{-0.216}	&	\textbf{-0.216}	&	\textbf{-0.205}	\\
	&	(0.019)	&	(0.019)	&	(0.066)	&	(0.019)	&	(0.020)	&	(0.019)	\\
\multirow{2}{*}{\quad 50s and older}	&	\textbf{-0.403}	&	\textbf{-0.375}	&	\textbf{-0.542}	&	\textbf{-0.403}	&	\textbf{-0.403}	&	\textbf{-0.392}	\\
	&	(0.017)	&	(0.017)	&	(0.054)	&	(0.017)	&	(0.016)	&	(0.016)	\\
Strata	&		&		&		&		&		&		\\
\multirow{2}{*}{\quad Medium}	&	\textbf{-0.049}	&	\textbf{-0.034}	&	-0.031	&	\textbf{-0.050}	&	\textbf{-0.050}	&	\textbf{-0.051}	\\
	&	(0.014)	&	(0.015)	&	(0.045)	&	(0.014)	&	(0.015)	&	(0.014)	\\
\multirow{2}{*}{\quad High}	&	\textbf{-0.051}	&	-0.032	&	-0.012	&	\textbf{-0.049}	&	\textbf{-0.050}	&	\textbf{-0.053}	\\
	&	(0.023)	&	(0.024)	&	(0.063)	&	(0.025)	&	(0.023)	&	(0.024)	\\
Risk perception marijuana use	&		&		&		&		&		&		\\
\multirow{2}{*}{\quad Medium}	&	\textbf{0.520}	&	\textbf{0.559}	&	\textbf{0.553}	&	\textbf{0.520}	&	\textbf{0.515}	&	\textbf{0.518}	\\
	&	(0.044)	&	(0.045)	&	(0.127)	&	(0.045)	&	(0.045)	&	(0.045)	\\
\multirow{2}{*}{\quad High}	&	\textbf{0.279}	&	\textbf{0.365}	&	\textbf{0.214}	&	\textbf{0.275}	&	\textbf{0.271}	&	\textbf{0.270}	\\
	&	(0.034)	&	(0.033)	&	(0.095)	&	(0.033)	&	(0.033)	&	(0.032)	\\
\multirow{2}{*}{Years of education}	&	\textbf{0.017}	&	\textbf{0.013}	&	\textbf{0.014}	&	\textbf{0.017}	&	\textbf{0.017}	&	\textbf{0.016}	\\
	&	(0.002)	&	(0.002)	&	(0.004)	&	(0.002)	&	(0.002)	&	(0.002)	\\
\multirow{2}{*}{Female}	&	\textbf{-0.310}	&	\textbf{-0.281}	&	\textbf{-0.471}	&	\textbf{-0.310}	&	\textbf{-0.312}	&	\textbf{-0.300}	\\
	&	(0.013)	&	(0.014)	&	(0.040)	&	(0.014)	&	(0.013)	&	(0.013)	\\
\multirow{2}{*}{Good mental health}	&	\textbf{-0.129}	&	\textbf{-0.113}	&	\textbf{-0.113}	&	\textbf{-0.128}	&	\textbf{-0.129}	&		\\
	&	(0.016)	&	(0.017)	&	(0.050)	&	(0.016)	&	(0.016)	&		\\
\multirow{2}{*}{Good physical health}	&	0.009	&	0.015	&	-0.017	&	0.010	&	0.009	&		\\
	&	(0.016)	&	(0.017)	&	(0.049)	&	(0.016)	&	(0.016)	&		\\
\multirow{2}{*}{Marijuana users in network}	&	\textbf{0.695}	&	\textbf{0.617}	&	\textbf{0.785}	&	\textbf{0.696}	&	\textbf{0.696}	&	\textbf{0.705}	\\
	&	(0.013)	&	(0.015)	&	(0.041)	&	(0.013)	&	(0.013)	&	(0.013)	\\
\multirow{2}{*}{Worker}	&	\textbf{0.127}	&	\textbf{0.123}	&	0.081	&	\textbf{0.127}	&	\textbf{0.127}	&	\textbf{0.121}	\\
	&	(0.014)	&	(0.013)	&	(0.047)	&	(0.014)	&	(0.014)	&	(0.014)	\\
Constant	&	\textbf{-0.527}	&	\textbf{-0.164}	&	-0.225	&	\textbf{-0.526}	&	\textbf{-0.518}	&	\textbf{-0.599}	\\
	&	(0.061)	&	(0.057)	&	(0.173)	&	(0.058)	&	(0.058)	&	(0.054)	\\
Regional-fixed effects	&	$\checkmark$	&	$\checkmark$	&	$\checkmark$	&	$\checkmark$	&	$\checkmark$	&	$\checkmark$	\\
\hline													
Sample size	&	49,414	&	49,414	&	5,574	&	49,414	&	49,414	&	49,414	\\
\hline											
\end{tabular}
				\begin{tablenotes}[para,flushleft]
	\footnotesize \textit{Notes}: Bold font indicates statistically significant variables. Robustness checks under different specifications and measures of relevant variables. Column (1) shows the baseline setting to facilitate comparisons.\\
	\textit{Source}: Authors' construction.
	\end{tablenotes}
	\end{minipage}}
\end{threeparttable}
\end{table}

%%%%%%% Posterior results of location parameters: Extensive equation %%%%%%%%%%%%%%
\begin{table}[h!]\centering \caption{Robustness checks: Posterior results of location parameters in the extensive margin of marijuana demand in Colombia.\label{tab:rob2}}
	\begin{threeparttable}
		\resizebox{1\textwidth}{!}{\begin{minipage}{\textwidth}
\begin{tabular}{l c c c c c c }\hline
\multirow{3}{*}{Variable}	&	Baseline	&	Access	&	Uni-personal	&	Interaction	&	Price	&	Perception	\\
	&	Specification	&	definition	&	household	&	Age and risk	&	definition	&	health status	\\
	&	(1)	&	(2)	&	(3)	&	(4)	&	(5)	&	(6)	\\
\hline													
Exclusionary restrictions	&		&		&		&		&		&		\\
\multirow{2}{*}{\quad  Alcohol and cigarette use}	&	\textbf{0.931}	&	\textbf{0.953}	&	\textbf{0.711}	&	\textbf{0.942}	&	\textbf{0.942}	&	\textbf{0.956}	\\
	&	(0.045)	&	(0.042)	&	(0.098)	&	(0.039)	&	'(0.042)	&	(0.042)	\\
Age	&		&		&		&		&		&		\\
\multirow{2}{*}{\quad 30s}	&	\textbf{-0.378}	&	\textbf{-0.385}	&	\textbf{-0.385}	&	\textbf{-0.385}	&	\textbf{-0.384}	&	\textbf{-0.379}	\\
	&	(0.044)	&	(0.047)	&	(0.117)	&	(0.044)	&	(0.043)	&	(0.044)	\\
\multirow{2}{*}{\quad 40s}	&	\textbf{-0.617}	&	\textbf{-0.613}	&	\textbf{-0.548}	&	\textbf{-0.603}	&	\textbf{-0.608}	&	\textbf{-0.591}	\\
	&	(0.054)	&	(0.061)	&	(0.134)	&	(0.055)	&	(0.060)	&	(0.058)	\\
\multirow{2}{*}{\quad 50s and older}	&	\textbf{-1.019}	&	\textbf{-1.041}	&	\textbf{-1.025}	&	\textbf{-1.026}	&	\textbf{-1.028}	&	\textbf{-1.015}	\\
	&	(0.066)	&	(0.063)	&	(0.131)	&	(0.064)	&	(0.063)	&	(0.063)	\\
Strata	&		&		&		&		&		&		\\
\multirow{2}{*}{\quad Medium}	&	\textbf{0.090}	&	\textbf{0.081}	&	0.082	&	\textbf{0.084}	&	\textbf{0.084}	&	0.077	\\
	&	(0.039)	&	(0.037)	&	(0.097)	&	(0.039)	&	(0.039)	&	(0.041)	\\
\multirow{2}{*}{\quad High}	&	\textbf{0.130}	&	\textbf{0.131}	&	-0.006	&	\textbf{0.133}	&	\textbf{0.132}	&	\textbf{0.132}	\\
	&	(0.063)	&	(0.06)	&	(0.128)	&	(0.058)	&	(0.057)	&	(0.061)	\\
Risk perception marijuana use	&		&		&		&		&		&		\\
\multirow{2}{*}{\quad Medium}	&	-0.112	&	-0.092	&	-0.043	&	-0.106	&	-0.105	&	-0.106	\\
	&	(0.081)	&	(0.073)	&	(0.176)	&	(0.076)	&	(0.080)	&	(0.077)	\\
\multirow{2}{*}{\quad High}	&	\textbf{-0.790}	&	\textbf{-0.773}	&	\textbf{-0.951}	&	\textbf{-0.786}	&	\textbf{-0.786}	&	\textbf{-0.788}	\\
	&	(0.072)	&	(0.067)	&	(0.160)	&	(0.067)	&	(0.071)	&	'(0.071)	\\
\multirow{2}{*}{Years of education}	&	\textbf{-0.012}	&	\textbf{-0.010}	&	-0.002	&	\textbf{-0.012}	&	\textbf{-0.012}	&	\textbf{-0.014}	\\
	&	(0.005)	&	(0.005)	&	(0.011)	&	(0.005)	&	(0.005)	&	(0.005)	\\
\multirow{2}{*}{Female}	&	\textbf{-0.449}	&	\textbf{-0.447}	&	\textbf{-0.344}	&	\textbf{-0.452}	&	\textbf{-0.452}	&	\textbf{-0.424}	\\
	&	(0.037)	&	(0.039)	&	(0.099)	&	(0.039)	&	(0.041)	&	(0.038)	\\
\multirow{2}{*}{Good mental health}	&	\textbf{-0.192}	&	\textbf{-0.192}	&	\textbf{-0.212}	&	\textbf{-0.189}	&	\textbf{-0.192}	&		\\
	&	(0.044)	&	(0.040)	&	(0.103)	&	(0.042)	&	(0.042)	&		\\
\multirow{2}{*}{Good physical health}	&	-0.033	&	-0.038	&	-0.061	&	-0.035	&	-0.034	&		\\
	&	(0.051)	&	(0.047)	&	(0.109)	&	(0.043)	&	(0.046)	&		\\
\multirow{2}{*}{Marijuana users in network}	&	\textbf{0.971}	&	\textbf{1.004}	&	\textbf{1.175}	&	\textbf{0.972}	&	\textbf{0.979}	&	\textbf{0.981}	\\
	&	(0.052)	&	(0.048)	&	(0.143)	&	(0.053)	&	(0.052)	&	(0.055)	\\
\multirow{2}{*}{Worker}	&	\textbf{-0.094}	&	\textbf{-0.093}	&	-0.166	&	\textbf{-0.100}	&	\textbf{-0.095}	&	\textbf{-0.115}	\\
	&	(0.037)	&	(0.037)	&	(0.101)	&	(0.039)	&	(0.039)	&	(0.037)	\\
Constant	&	\textbf{-1.675}	&	\textbf{-1.813}	&	\textbf{-1.516}	&	\textbf{-1.681}	&	\textbf{-1.694}	&	\textbf{-1.830}	\\
	&	(0.159)	&	(0.163)	&	0.465	&	(0.167)	&	(0.177)	&	(0.162)	\\
Regional-fixed effects	&	$\checkmark$	&	$\checkmark$	&	$\checkmark$	&	$\checkmark$	&	$\checkmark$	&	$\checkmark$	\\
\hline													
Sample size	&	49,414	&	49,414	&	5,574	&	49,414	&	49,414	&	49,414	\\
\hline													
\end{tabular}
				\begin{tablenotes}[para,flushleft]
	\footnotesize \textit{Notes}: Bold font indicates statistically significant variables. Robustness checks under different specifications and measures of relevant variables. Column (1) shows the baseline setting to facilitate comparisons. \\
	\textit{Source}: Authors' construction.
	\end{tablenotes}
	\end{minipage}}
\end{threeparttable}
\end{table}

%%%%%%% Posterior results of location parameters: Intensive equation %%%%%%%%%%%%%%
\begin{table}[h!]\centering \caption{Robustness checks: Posterior results of location parameters in the intensive margin of marijuana demand in Colombia.\label{tab:rob3}}
	\begin{threeparttable}
		\resizebox{1\textwidth}{!}{\begin{minipage}{\textwidth}
\begin{tabular}{l c c c c c c }\hline
\multirow{3}{*}{Variable}	&	Baseline	&	Access	&	Uni-personal	&	Interaction	&	Price	&	Perception	\\
	&	Specification	&	definition	&	household	&	Age and risk	&	definition	&	health status	\\
	&	(1)	&	(2)	&	(3)	&	(4)	&	(5)	&	(6)	\\
\hline													
Age	&		&		&		&		&		&		\\
\multirow{2}{*}{\quad 30s}	&	0.457	&	0.593	&	3.192	&	0.859	&	0.562	&	0.648	\\
	&	(1.302)	&	(1.278)	&	(2.951)	&	(1.391)	&	(1.962)	&	(1.313)	\\
\multirow{2}{*}{\quad 40s}	&	0.534	&	0.518	&	0.521	&	-0.041	&	-2.732	&	0.633	\\
	&	(1.899)	&	(1.941)	&	(3.785)	&	(1.974)	&	(3.111)	&	(2.008)	\\
\multirow{2}{*}{\quad 50s and older}	&	-2.995	&	-2.449	&	-5.609	&	-2.568	&	1.585	&	-2.384	\\
	&	(2.197)	&	(2.090)	&	(4.361)	&	(2.360)	&	(2.675)	&	(2.128)	\\
Strata	&		&		&		&		&		&		\\
\multirow{2}{*}{\quad Medium}	&	-0.067	&	-0.067	&	0.116	&	-0.074	&	-0.112	&	-0.084	\\
	&	(0.099)	&	(0.091)	&	(0.224)	&	(0.091)	&	(0.093)	&	(0.091)	\\
\multirow{2}{*}{\quad High}	&	-0.199	&	-0.168	&	-0.340	&	-0.208	&	-0.149	&	-0.214	\\
	&	(0.149)	&	(0.132)	&	(0.280)	&	(0.135)	&	(0.129)	&	(0.127)	\\
Risk perception marijuana use	&		&		&		&		&		&		\\
\multirow{2}{*}{\quad Medium}	&	-0.204	&	-0.040	&	-0.701	&	-0.144	&	-0.064	&	-0.036	\\
	&	(0.300)	&	'(0.147)	&	(0.459)	&	(0.301)	&	(0.161)	&	(0.152)	\\
\multirow{2}{*}{\quad Age 30s $\times$ Medium}	&		&		&		&	-0.328	&		&		\\
	&		&		&		&	(0.384)	&		&		\\
\multirow{2}{*}{\quad Age 40s $\times$ Medium}	&		&		&		&	0.544	&		&		\\
	&		&		&		&	(0.628)	&		&		\\
\multirow{2}{*}{\quad Age 50s and older $\times$ Medium}	&		&		&		&	-0.516	&		&		\\
	&		&		&		&	(0.674)	&		&		\\
\multirow{2}{*}{\quad High}	&	-0.327	&	-0.162	&	\textbf{-0.818}	&	-0.302	&	-0.208	&	-0.201	\\
	&	(0.247)	&	(0.137)	&	(0.413)	&	(0.256)	&	(0.142)	&	(0.134)	\\
\multirow{2}{*}{\quad Age 30s $\times$ High}	&		&		&		&	-0.319	&		&		\\
	&		&		&		&	(0.355)	&		&		\\
\multirow{2}{*}{\quad Age 40s $\times$ High}	&		&		&		&	0.558	&		&		\\
	&		&		&		&	(0.550)	&		&		\\
\multirow{2}{*}{\quad Age 50s and older $\times$ High}	&		&		&		&	-0.273	&		&		\\
	&		&		&		&	(0.551)	&		&		\\
\multirow{2}{*}{Years of education}	&	\textbf{-0.069}	&	\textbf{-0.065}	&	\textbf{-0.081}	&	\textbf{-0.068}	&	\textbf{-0.060}	&	\textbf{-0.065}	\\
	&	(0.015)	&	(0.013)	&	(0.031)	&	(0.016)	&	(0.012)	&	(0.013)	\\
\multirow{2}{*}{Female}	&	\textbf{-0.607}	&	\textbf{-0.582}	&	-0.421	&	\textbf{-0.599}	&	\textbf{-0.640}	&	\textbf{-0.627}	\\
	&	(0.146)	&	(0.097)	&	(0.336)	&	(0.115)	&	(0.100)	&	(0.094)	\\
\multirow{2}{*}{Good mental health}	&	0.052	&	0.050	&	0.201	&	0.057	&	-0.005	&		\\
	&	(0.107)	&	(0.093)	&	(0.279)	&	(0.103)	&	(0.094)	&		\\
\multirow{2}{*}{Good physical health}	&	-0.110	&	-0.115	&	-0.304	&	-0.128	&	-0.115	&		\\
	&	(0.105)	&	(0.110)	&	(0.277)	&	(0.112)	&	(0.108)	&		\\
\multirow{2}{*}{Marijuana users in network}	&	\textbf{0.619}	&	\textbf{0.652}	&	0.364	&	\textbf{0.663}	&	\textbf{0.844}	&	\textbf{0.827}	\\
	&	(0.306)	&	(0.144)	&	(0.606)	&	(0.271)	&	(0.151)	&	(0.157)	\\
\multirow{2}{*}{Worker}	&	0.151	&	0.153	&	0.140	&	0.143	&	0.135	&	0.157	\\
	&	(0.096)	&	(0.092)	&	(0.233)	&	(0.088)	&	(0.085)	&	(0.086)	\\
Price 	&		&		&		&		&		&		\\
\multirow{2}{*}{\quad $\log\left\{\text{price of marijuana}\right\}$}	&	\textbf{-0.445}	&	\textbf{-0.436}	&	-0.312	&	\textbf{-0.463}	&	\textbf{-0.363}	&	\textbf{-0.437}	\\
	&	(0.101)	&	(0.097)	&	(0.251)	&	(0.096)	&	(0.128)	&	(0.092)	\\
\multirow{2}{*}{\quad Age 30s $\times \log\left\{\text{price of marijuana}\right\}$}	&	-0.033	&	-0.058	&	-0.325	&	-0.050	&	-0.049	&	-0.063	\\
	&	(0.167)	&	(0.166)	&	(0.376)	&	(0.174)	&	(0.249)	&	(0.171)	\\
\multirow{2}{*}{\quad Age 40s $\times \log\left\{\text{price of marijuana}\right\}$}	&	-0.046	&	-0.055	&	0.030	&	-0.039	&	0.370	&	-0.065	\\
	&	(0.245)	&	(0.252)	&	(0.485)	&	(0.244)	&	(0.400)	&	-0.261	\\
\multirow{2}{*}{\quad Age 50s and older $\times \log\left\{\text{price of marijuana}\right\}$}	&	0.337	&	0.258	&	0.653	&	0.309	&	-0.269	&	0.252	\\
	&	(0.283)	&	(0.265)	&	(0.553)	&	(0.291)	&	(0.343)	&	-0.269	\\
Constant	&	\textbf{5.043}	&	\textbf{4.798}	&	\textbf{4.899}	&	\textbf{5.107}	&	\textbf{3.666}	&	\textbf{4.411}	\\
	&	(1.352)	&	(0.884)	&	(2.374)	&	(1.196)	&	(1.119)	&	(0.836)	\\
Regional-fixed effects	&	$\checkmark$	&	$\checkmark$	&	$\checkmark$	&	$\checkmark$	&	$\checkmark$	&	$\checkmark$	\\
\hline													
Sample size	&	49,414	&		&	5,574	&	49,414	&	49,414	&	49,414	\\
\hline													
													
\end{tabular}
				\begin{tablenotes}[para,flushleft]
	\footnotesize \textit{Notes}: Bold font indicates statistically significant variables. Robustness checks under different specifications and measures of relevant variables. Column (1) shows the baseline setting to facilitate comparisons.\\
	\textit{Source}: Authors' construction.
	\end{tablenotes}
	\end{minipage}}
\end{threeparttable}
\end{table}

In general, it seems that the results of the baseline specification are robust, there are not statistically significant differences in most of the cases compared to the alternative measures of relevant variables or model specifications. We observe that the same variables are statistically significant in all three stages of demand for marijuana, and the numeric values of the posterior estimates are relatively similar. However, the sub-sample of uni-personal households present some intriguing results, potentially due to power issues.

\section{Policy analysis}\label{policy}

\subsection{Marijuana legalization}

We perform some counterfactual experiments to estimate the potential effects of the legalization of marijuana in Colombia for different representative individuals. Particularly, we estimate the posterior predictive probability for individual 0,
\begin{align*}    p(A_0,C_0,Y_0|\bm{T},\bm{X})&=\int_{\bm{S}}\int_{\boldsymbol{\Theta}} \left\{\mathds{1}(A_{0}=0)\mathds{1}(U_{0}^a\leq 0)+\mathds{1}(A_{0}=1)\mathds{1}(U_{0}^a > 0)\right.\\
	&\left.\times\left[\mathds{1}(C_{0}=0)\mathds{1}(U_{0}^c\leq 0)+\mathds{1}(C_{0}=1)\mathds{1}(C_{0}^c> 0)\right]\right\}\\
 &\times p(A_0|\bm{T},\bm{X})\times p(C_0|A_0,\bm{T},\bm{X})\times p(Y_0|C_0,A_0,\bm{T},\bm{X})  \pi(\boldsymbol{\theta},\boldsymbol{\Sigma}|\bm{T},\bm{X})d\boldsymbol{\theta}d\boldsymbol{\Sigma}, 
\end{align*}

where $\bm{S}$ is the support of integration of $\boldsymbol{\Sigma}$, $p(A_0=1|\bm{T},\bm{X})=1-\Phi(\bm{x}_0^{a\top}\boldsymbol{\theta}^a,1)$, $p(C_0=1|A_0=1,\bm{T},\bm{X})=1-\Phi(\bm{x}_0^{c\top}\boldsymbol{\theta}^c+\sigma_{ac}(U_0^a-\bm{x}_0^{a\top}\boldsymbol{\theta}^a),1-\sigma_{ac}^2)$ and $p(Y_0|C_0=1,A_0=1,\bm{T},\bm{X})\sim N(\mu_{y|ac},\sigma^2_{y|ac})$, where \begin{align*}
    \mu_{y|ac}&=\bm{x}_0^{y\top}\boldsymbol{\theta}^y+[\sigma_{ya} \ \sigma_{yc}]\begin{bmatrix}
    1 & \sigma_{ac}\\
    \sigma_{ac} & 1
\end{bmatrix}^{-1}\begin{bmatrix}
    U_0^a-\bm{x}_0^{a\top}\boldsymbol{\theta}^a\\
    U_0^c-\bm{x}_0^{c\top}\boldsymbol{\theta}^c
\end{bmatrix},
\end{align*}
and 
\begin{align*} 
\sigma^2_{y|ac}&=\sigma^2_{y}-[\sigma_{ya} \ \sigma_{yc}]\begin{bmatrix}
    1 & \sigma_{ac}\\
    \sigma_{ac} & 1
\end{bmatrix}^{-1}\begin{bmatrix}
    \sigma_{ya}\\
    \sigma_{yc}
\end{bmatrix}. 
\end{align*}
Observe the relevance of the selection parameters, $\sigma_{ac}$, $\sigma_{ya}$ and $\sigma_{yc}$ in the previous expressions. These account for unobserved dependence between the three stages of the demand for marijuana.

The above integral can be estimated in a straight forward way using the draws from the posterior distribution. Therefore, we use simulation to estimate the effects of the legalization of marijuana on the probability of use, given that under legalization the probability of access is equal to 1, that is, $p(A_0=1|\bm{T},\bm{X})=1$, and the amount of consumption, conditional on use, where we take into account that we model $\log(Y_{it})$, so we get by simulation $Y_{it}$, that is, the amount of joints per month.  

We show in tables \ref{tab:legal1} and \ref{tab:legal2} the results of these exercises for the representative individual who has access to marijuana. In particular, this is an individual with 12 years of education, working, good self-perception of health status, family members and friends who do not consume marijuana, but she/he has consumed alcohol and cigarettes, lives in Medellín in a low socioeconomic stratum, and there is a drug dealer in the neighborhood. We have in these tables seven scenarios, rows one to three in each panel show results under different scenarios about risk perceptions of marijuana use, these experiments are based on the fact that marketing campaigns warning about bad consequences of consuming some products may curve demand. For instance, warning labeling explains why consumers become more health-conscious, and consequently, more risk-averse \citep{barahona2023, Berg2023, cannoy2023, Kaai2023,Nguyen2023,Nian2023,brennan2022}. Rows four to seven show results under different price scenarios, the ones that we analyze in the next subsection for potential tax revenues, taxes contribute to raise revenues for public policy, and help to curve the demand function \citep{allcott2019}.

We can see in Table \ref{tab:legal1} the results for the representative woman, where each panel shows results by age spline. For instance, the first row in the first panel shows the results under a the baseline price for this representative woman (US\cent 78.2), who has a high risk perception about marijuana use. We observe that the predicted probability of having access to marijuana is 74.5\%, and the predicted probabilities of marijuana use are 1.30\% and 1.75\%, overall women with these features, and those with access, respectively. This means that the probability of use overall these representative women increases 0.44 percentage points given a policy of legalization of marijuana. Conditional on access and use, the predicted consumption for this representative woman is 5.9 joints per month. All these estimates has the standard errors that are calculated by simulation using repeated sampling.

We observe from Table \ref{tab:legal1} that under legalization of marijuana, risk perception has a higher effect among women in decreasing the probability of use than price. For instance, the second and third rows show that given access, the probability of use increases 9.37 and 4.58 percentage points for women who have a medium and low risk perceptions about marijuana use, compared with just 0.44 p.p. for women with high risk perception. However, the effect of risk perception decreases with age. We observe the same pattern among men (see Table \ref{tab:legal2}). Overall, taking into account these three risk scenarios, age splines and gender, and using the expansion factors of the survey, we find that the probability of use increases from 2.3\% pre-legalization to 3.0\% under legalization.

Given the effect of risk perception, and the patterns of this by age splines, we can deduce that marketing campaigns targeting young individuals will be an effective way to curve demand for marijuana under a legal setting. First, young have a lower risk perception about marijuana use than other age splines, the sample average for a high risk is 88\% for the former, whereas this percentage is equal to 92\%, 93\% and 94\% for 30s, 40s and 50s age splines, this means that there is a higher gap among young individuals. And second, reducing more the demand of this group implies that the cumulative effect on consumption through their life span is higher, with potentially more good externalities. However, we should take into account that the effect of risk perception has a limit, this is, on average 91\% of the individuals already have a high risk perception about marijuana use. This fact motivates to perform experiments with different prices (taxes), which also means different scenarios regarding government revenues.

Thus, the second set of experiments consider the effect of legalization of marijuana on price, and as a consequence, on access, extensive and intensive margins. Particularly, marijuana legalization implies that the inherent extra cost due to illegality would disappear. However, we should take a potential tax into account. A first benchmark is the average cost of marijuana in Colombia, \cite{velez2021medicinal} found that the average production cost is US\cent 1 per gram, and taking into account that the average percentage of distribution cost in Colombia is 15\%, the cost of one joint of marijuana is approximately US\cent 1.15. In addition, the average return of capital in Colombia is around 15.25\% according to Corficolombiana, a prestigious financial institution in this country, fluctuating between 13.9\% y 16.6\%,\footnote{See \href{chrome-extension://efaidnbmnnnibpcajpcglclefindmkaj/https://investigaciones.corficolombiana.com/documents}{Rentabilidad esperada del capital propio.}} thus a gram of marijuana has a base price without taxes of approximately US\cent 1.33. The first counterfactual exercise uses as reference the tax on cigarettes, which is US\cent 5.9 in Colombia,\footnote{Real price 2019, see \href{https://actualicese.com/certificacion-04-del-07-12-2021/}{CERTIFICACIÓN 4, 2021 of Dirección General de Apoyo Fiscal del Ministerio de Hacienda.}} then the potential price of one gram of marijuana, tax included, is approximately US\cent 7.3. This price can be considered as a potential lower bound due to being less expensive that the legal price of cigarettes in Colombia, which is on average US\cent 11.5. Although, this price is higher than the price of illegal cigarettes, US\cent 5.3.\footnote{Real price 2019, see \href{https://www.semana.com/salud/articulo/consumo-de-cigarrillos-34-de-cada-100-fueron-de-contrabando-esto-revelo-la-federacion-nacional-de-departamentos/202315/}{Estudio de incidencia del consumo de cigarrillos en colombia 2022.}} The second exercise assumes that the price of marijuana is equal to the legal price of cigarettes (US\cent 11.5). Finally, the potential lower bound and the actual price of marijuana offer a spectrum of possibilities for tax scenarios. Thus, we perform two more experiments, 50\% decrease and 25\% increase with respect to the actual price of marijuana. The former implies a tax of US\cent 37.8 per joint, which is more than 6 times the tax on cigarettes. The latter is based on \cite{jacobi2016marijuana}, who propose a 25\% tax on the actual price of marijuana in Australia. In the Colombian case, this tax would be more than 16 times the tax of cigarettes. We think about the this scenario as a potential upper bound price in the case of legalization of marijuana due to a relatively high price may imply a huge black market of marijuana in this country. For instance, the size of the black market of cigarettes in Colombia is 34\%.\footnote{See \href{https://www.semana.com/salud/articulo/consumo-de-cigarrillos-34-de-cada-100-fueron-de-contrabando-esto-revelo-la-federacion-nacional-de-departamentos/202315/}{Estudio de incidencia del consumo de cigarrillos en colombia 2022.}} 

The fourth to seventh rows in each panel of tables \ref{tab:legal1} and \ref{tab:legal2} show the results. We observe that there are not remarkable differences regarding the probability of access and use under different price settings; however, the intensity of consumption decreases with price, as expected. The shape of the intensity margin as a function of age splines has an ``inverted U-shape", that is, this is low for 20s and 50s, increases in 30s, and has a peak in 40s. This latter group has a very high level of consumption, however, we should take with caution this result due to the also high volatility level. Observe that price helps to curve the intensity of consumption under a legalization policy; however, there are not significant changes in the extensive margin due to different price regimes.  

Other patterns that we observe from tables \ref{tab:legal1} and \ref{tab:legal2} are that access is lower for individuals who have a low risk perception regarding marijuana use; however, the probability of marijuana use is higher for this group. In addition, access also decreases with age, and is higher for men, who in turn have a higher probability of use, and given use, have a substantially higher level of consumption, approximately two times the level of women. 

We perform \textit{ceteris paribus} exercises in order to isolate the effects of different control variables, and get a better understanding of the situation. However, \cite{manthey2023} demonstrated that warning information about product use that potentially curve consumption must be complemented with taxes (pricing). Both approaches should be integrated as part of a comprehensive strategy aimed at mitigating the burdens on public health, social well-being, and economy, of marijuana use.

\begin{sidewaystable}
%\begin{table}[h!]    
\centering \caption{Counterfactual experiments: Effects of legalization of marijuana in Colombia for a representative woman.\label{tab:legal1}}
	\begin{threeparttable}
		\resizebox{1\textwidth}{!}{\begin{minipage}{\textwidth}
				\begin{tabular}{l c c c c c c c c c c c c c c c c}\hline
	&		&		&	&	\multicolumn{2}{c}{Pred. Prob. access}			&	&	\multicolumn{7}{c}{Pred. prob use}											&	&	\multicolumn{2}{c}{Pred. consumption}			\\
							\cline{5-6}					\cline{8-14}													\cline{16-17}			
		\multicolumn{3}{c}{Scenario}			&	&	\multicolumn{2}{c}{All}			&	&	\multicolumn{2}{c}{All}			&	&	\multicolumn{2}{c}{Access}			&	&	Change	&	&	\multicolumn{2}{c}{Consumption}			\\
\cline{1-3}							\cline{5-6}					\cline{8-9}					\cline{11-12}					\cline{14-14}			\cline{16-17}			
Price scenario	&	Price	&	Risk perception	&	&	Mean	&	Std. Error	&	&	Mean	&	Std. Error	&	&	Mean	&	Std. Error	&	&	Mean	&	&	Mean	&	Std. Error	\\
		\hline																										
		\multicolumn{16}{c}{Woman in 20s or younger}																										\\
		\hline																										
Baseline	&	US\cent 78.2	&	High	&	&	74.50\%	&	0.0137	&	&	1.30\%	&	0.0036	&	&	1.74\%	&	0.0048	&	&	0.44 p.p.	&	&	5.90	&	0.30	\\
Baseline	&	US\cent 78.2	&	Medium	&	&	77.70\%	&	0.0131	&	&	8.30\%	&	0.0087	&	&	10.68\%	&	0.0110	&	&	9.37 p.p.	&	&	6.58	&	0.33	\\
Baseline	&	US\cent 78.2	&	Low	&	&	66.50\%	&	0.0149	&	&	9.10\%	&	0.0091	&	&	13.68\%	&	0.0130	&	&	4.58 p.p.	&	&	9.06	&	0.50	\\
Lower bound	&	US\cent 7.3	&	High	&	&	70.80\%	&	0.0143	&	&	2.30\%	&	0.0047	&	&	3.24\%	&	0.0067	&	&	0.94 p.p.	&	&	17.44	&	0.87	\\
Cigarette	&	US\cent 11.5	&	High	&	&	72.70\%	&	0.0142	&	&	1.50\%	&	0.0038	&	&	2.08\%	&	0.0053	&	&	0.58 p.p.	&	&	13.74	&	0.64	\\
50\% decrease	&	US\cent 39.1	&	High	&	&	72.60\%	&	0.0141	&	&	1.60\%	&	0.0040	&	&	2.20\%	&	0.0054	&	&	0.60 p.p.	&	&	8.22	&	0.39	\\
25\% increase	&	US\cent 97.8	&	High	&	&	72.00\%	&	0.0141	&	&	2.60\%	&	0.0050	&	&	3.58\%	&	0.0069	&	&	0.98 p.p.	&	&	5.18	&	0.31	\\
		\hline																										
		\multicolumn{16}{c}{Woman in 30s or younger}																										\\
		\hline																										
Baseline	&	US\cent 78.2	&	High	&	&	69.40\%	&	0.0146	&	&	0.50\%	&	0.0022	&	&	0.72\%	&	0.0032	&	&	0.22 p.p.	&	&	22.20	&	2.64	\\
Baseline	&	US\cent 78.2	&	Medium	&	&	76.70\%	&	0.0133	&	&	2.50\%	&	0.0049	&	&	3.25\%	&	0.0064	&	&	0.75 p.p.	&	&	20.80	&	1.65	\\
Baseline	&	US\cent 78.2	&	Low	&	&	62.80\%	&	0.0152	&	&	3.80\%	&	0.0061	&	&	6.05\%	&	0.0095	&	&	2.25 p.p.	&	&	30.12	&	2.80	\\
Lower bound	&	US\cent 7.3	&	High	&	&	70.50\%	&	0.0144	&	&	0.01\%	&	0.0030	&	&	1.28\%	&	0.1123	&	&	1.27 p.p.	&	&	48.21	&	3.87	\\
Cigarette	&	US\cent 11.5	&	High	&	&	67.50\%	&	0.0148	&	&	0.05\%	&	0.0022	&	&	0.74\%	&	0.0033	&	&	0.69 p.p.	&	&	47.54	&	7.40	\\
50\% decrease	&	US\cent 39.1	&	High	&	&	71.80\%	&	0.0142	&	&	0.40\%	&	0.0020	&	&	0.55\%	&	0.0028	&	&	0.15 p.p.	&	&	23.29	&	1.87	\\
25\% increase	&	US\cent 97.8	&	High	&	&	65.50\%	&	0.0150	&	&	0.40\%	&	0.0019	&	&	0.61\%	&	0.0030	&	&	0.21 p.p.	&	&	17.72	&	1.62	\\
		\hline																										
		\multicolumn{16}{c}{Woman in 40s or younger}																										\\
		\hline																										
Baseline	&	US\cent 78.2	&	High	&	&	67.10\%	&	0.0150	&	&	0.30\%	&	0.0013	&	&	0.44\%	&	0.0025	&	&	0.14 p.p.	&	&	57.68	&	10.86	\\
Baseline	&	US\cent 78.2	&	Medium	&	&	73.00\%	&	0.0140	&	&	2.00\%	&	0.0044	&	&	2.74\%	&	0.0060	&	&	0.74 p.p.	&	&	69.14	&	15.44	\\
Baseline	&	US\cent 78.2	&	Low	&	&	60.80\%	&	0.0154	&	&	1.90\%	&	0.0043	&	&	3.12\%	&	0.0070	&	&	1.24 p.p.	&	&	133.25	&	34.22	\\
Lower bound	&	US\cent 7.3	&	High	&	&	66.30\%	&	0.0150	&	&	0.30\%	&	0.0017	&	&	0.45\%	&	0.0026	&	&	0.15 p.p.	&	&	313.13	&	110.55	\\
Cigarette	&	US\cent 11.5	&	High	&	&	62.20\%	&	0.0153	&	&	0.60\%	&	0.0024	&	&	0.96\%	&	0.0039	&	&	0.36 p.p.	&	&	172.62	&	40.93	\\
50\% decrease	&	US\cent 39.1	&	High	&	&	64.30\%	&	0.0151	&	&	0.20\%	&	0.0014	&	&	0.31\%	&	0.0022	&	&	0.08 p.p.	&	&	76.03	&	16.98	\\
25\% increase	&	US\cent 97.8	&	High	&	&	68.90\%	&	0.0146	&	&	0.20\%	&	0.0014	&	&	0.29\%	&	0.0021	&	&	0.09 p.p.	&	&	63.72	&	11.05	\\
		\hline																										
		\multicolumn{16}{c}{Woman in 50s or younger}																										\\
		\hline																										
Baseline	&	US\cent 78.2	&	High	&	&	64.00\%	&	0.0152	&	&	0.10\%	&	0.0010	&	&	0.16\%	&	0.0015	&	&	0.06 p.p.	&	&	1.52	&	0.17	\\
Baseline	&	US\cent 78.2	&	Medium	&	&	68.50\%	&	0.0147	&	&	0.90\%	&	0.0020	&	&	1.31\%	&	0.0043	&	&	0.41 p.p.	&	&	1.72	&	0.22	\\
Baseline	&	US\cent 78.2	&	Low	&	&	56.70\%	&	0.0157	&	&	0.40\%	&	0.0020	&	&	0.70\%	&	0.0035	&	&	0.30 p.p.	&	&	2.47	&	0.32	\\
Lower bound	&	US\cent 7.3	&	High	&	&	59.90\%	&	0.0155	&	&	0.00\%	&	0.0000	&	&	0.00\%	&	0.0000	&	&	0.00 p.p.	&	&	5.40$^*$	&	1.27	\\
Cigarette	&	US\cent 11.5	&	High	&	&	60.30\%	&	0.0155	&	&	0.30\%	&	0.0017	&	&	0.50\%	&	0.0029	&	&	0.20 p.p.	&	&	6.18	&	1.64	\\
50\% decrease	&	US\cent 39.1	&	High	&	&	61.90\%	&	0.0154	&	&	0.00\%	&	0.0000	&	&	0.00\%	&	0.0000	&	&	0.00 p.p.	&	&	2.51$^*$	&	0.50	\\
25\% increase	&	US\cent 97.8	&	High	&	&	60.70\%	&	0.0155	&	&	0.20\%	&	0.0014	&	&	0.33\%	&	0.0023	&	&	0.13 p.p.	&	&	1.63	&	0.29	\\

\hline																																		\end{tabular}
				\begin{tablenotes}[para,flushleft]
	\footnotesize \textit{Notes}: $^*$ Up to four digits the probability of marijuana use is zero for this individual, but in the highly unlikely case of consumption, this is the expected amount. Predicted measures for the representative woman with access to marijuana. Her level of education is high school (12 years of education), works, has a good self-perception of mental and physical health, no marijuana users in her network, consumes alcohol and cigarettes, lives in Medellín in a low socioeconomic stratum, and there is a drug dealer in her neighborhood. Pred. stands for prediction, and prob. is probability.\\	
	\textit{Source}: Authors' construction.
	\end{tablenotes}
	\end{minipage}}
\end{threeparttable}
%\end{table}
\end{sidewaystable}

\begin{sidewaystable}
%\begin{table}[h!]    
\centering \caption{Counterfactual experiments: Effects of legalization of marijuana in Colombia for a representative man.\label{tab:legal2}}
	\begin{threeparttable}
		\resizebox{1\textwidth}{!}{\begin{minipage}{\textwidth}
				\begin{tabular}{l c c c c c c c c c c c c c c c c}\hline
	&		&		&	&	\multicolumn{2}{c}{Pred. Prob. access}			&	&	\multicolumn{7}{c}{Pred. prob use}											&	&	\multicolumn{2}{c}{Pred. consumption}			\\
							\cline{5-6}					\cline{8-14}													\cline{16-17}			
		\multicolumn{3}{c}{Scenario}			&	&	\multicolumn{2}{c}{All}			&	&	\multicolumn{2}{c}{All}			&	&	\multicolumn{2}{c}{Access}			&	&	Change	&	&	\multicolumn{2}{c}{Consumption}			\\
\cline{1-3}							\cline{5-6}					\cline{8-9}					\cline{11-12}					\cline{14-14}			\cline{16-17}			
Price scenario	&	Price	&	Risk perception	&	&	Mean	&	Std. Error	&	&	Mean	&	Std. Error	&	&	Mean	&	Std. Error	&	&	Mean	&	&	Mean	&	Std. Error	\\
		\hline																										
		\multicolumn{16}{c}{Man in 20s or younger}																										\\
		\hline																										
Baseline	&	US\cent 78.2	&	High	&	&	79.30\%	&	0.0128	&	&	4.90\%	&	0.0068	&	&	6.10\%	&	0.0085	&	&	1.30 p.p.	&	&	10.45	&	0.55	\\
Baseline	&	US\cent 78.2	&	Medium	&	&	83.70\%	&	0.0117	&	&	17.80\%	&	0.0121	&	&	21.26\%	&	0.0142	&	&	3.46 p.p.	&	&	10.44	&	0.55	\\
Baseline	&	US\cent 78.2	&	Low	&	&	71.40\%	&	0.0143	&	&	15.30\%	&	0.0114	&	&	21.43\%	&	0.0154	&	&	6.13 p.p.	&	&	15.17	&	0.76	\\
Lower bound	&	US\cent 7.3	&	High	&	&	80.40\%	&	0.0126	&	&	5.60\%	&	0.0073	&	&	6.96\%	&	0.0089	&	&	1.36 p.p.	&	&	29.38	&	1.47	\\
Cigarette	&	US\cent 11.5	&	High	&	&	77.50\%	&	0.0132	&	&	5.20\%	&	0.007	&	&	6.70\%	&	0.0090	&	&	1.50 p.p.	&	&	25.29	&	1.41	\\
50\% decrease	&	US\cent 39.1	&	High	&	&	72.20\%	&	0.0132	&	&	4.80\%	&	0.0067	&	&	6.21\%	&	0.0087	&	&	0.60 p.p.	&	&	13.88	&	0.62	\\
25\% increase	&	US\cent 97.8	&	High	&	&	79.60\%	&	0.0127	&	&	4.40\%	&	0.0065	&	&	5.53\%	&	0.0081	&	&	1.13 p.p.	&	&	9.98	&	0.46	\\
		\hline																										
		\multicolumn{16}{c}{Man in 30s or younger}																										\\
		\hline																										
Baseline	&	US\cent 78.2	&	High	&	&	76.70\%	&	0.0133	&	&	2.70\%	&	0.0051	&	&	3.52\%	&	0.0066	&	&	0.82 p.p.	&	&	34.46	&	3.09	\\
Baseline	&	US\cent 78.2	&	Medium	&	&	82.30\%	&	0.0120	&	&	9.50\%	&	0.0092	&	&	11.54\%	&	0.0111	&	&	2.04 p.p.	&	&	40.53	&	4.50	\\
Baseline	&	US\cent 78.2	&	Low	&	&	71.70\%	&	0.0142	&	&	6.80\%	&	0.0079	&	&	9.48\%	&	0.0109	&	&	2.68 p.p.	&	&	48.52	&	4.64	\\
Lower bound	&	US\cent 7.3	&	High	&	&	76.40\%	&	0.0134	&	&	2.20\%	&	0.0046	&	&	2.88\%	&	0.0061	&	&	0.68 p.p.	&	&	94.66	&	8.77	\\
Cigarette	&	US\cent 11.5	&	High	&	&	77.90\%	&	0.0131	&	&	2.30\%	&	0.0047	&	&	2.95\%	&	0.0061	&	&	0.65 p.p.	&	&	74.21	&	6.24	\\
50\% decrease	&	US\cent 39.1	&	High	&	&	77.90\%	&	0.0131	&	&	1.90\%	&	0.0043	&	&	2.44\%	&	0.0055	&	&	0.54 p.p.	&	&	44.01	&	3.63	\\
25\% increase	&	US\cent 97.8	&	High	&	&	74.70\%	&	0.0137	&	&	1.00\%	&	0.0031	&	&	1.34\%	&	0.0042	&	&	0.34 p.p.	&	&	33.54	&	4.44	\\
		\hline																										
		\multicolumn{16}{c}{Man in 40s or younger}																										\\
		\hline																										
Baseline	&	US\cent 78.2	&	High	&	&	75.20\%	&	0.0137	&	&	0.70\%	&	0.0026	&	&	0.93\%	&	0.0035	&	&	0.23 p.p.	&	&	123.18	&	43.60	\\
Baseline	&	US\cent 78.2	&	Medium	&	&	80.20\%	&	0.0126	&	&	4.20\%	&	0.0063	&	&	5.23\%	&	0.0078	&	&	1.03 p.p.	&	&	106.25	&	26.12	\\
Baseline	&	US\cent 78.2	&	Low	&	&	67.00\%	&	0.0148	&	&	6.20\%	&	0.0076	&	&	9.25\%	&	0.0112	&	&	3.05 p.p.	&	&	197.54	&	65.80	\\
Lower bound	&	US\cent 7.3	&	High	&	&	74.80\%	&	0.0137	&	&	1.20\%	&	0.0034	&	&	1.60\%	&	0.0045	&	&	0.40 p.p.	&	&	405.79	&	128.76	\\
Cigarette	&	US\cent 11.5	&	High	&	&	75.10\%	&	0.0136	&	&	0.90\%	&	0.0030	&	&	1.99\%	&	0.0039	&	&	1.09 p.p.	&	&	266.60	&	67.18	\\
50\% decrease	&	US\cent 39.1	&	High	&	&	73.60\%	&	0.0139	&	&	1.10\%	&	0.0033	&	&	1.49\%	&	0.0044	&	&	0.39 p.p.	&	&	272.48	&	135.76	\\
25\% increase	&	US\cent 97.8	&	High	&	&	72.20\%	&	0.0141	&	&	1.70\%	&	0.0040	&	&	2.35\%	&	0.0056	&	&	0.65 p.p.	&	&	109.08	&	19.89	\\
		\hline																										
		\multicolumn{16}{c}{Man in 50s or younger}																										\\
		\hline																										
Baseline	&	US\cent 78.2	&	High	&	&	70.30\%	&	0.0144	&	&	0.40\%	&	0.0019	&	&	0.57\%	&	0.0028	&	&	0.17 p.p.	&	&	3.13	&	0.43	\\
Baseline	&	US\cent 78.2	&	Medium	&	&	73.80\%	&	0.0139	&	&	1.60\%	&	0.0039	&	&	2.17\%	&	0.0053	&	&	0.57 p.p.	&	&	4.21	&	0.73	\\
Baseline	&	US\cent 78.2	&	Low	&	&	61.00\%	&	0.0154	&	&	1.70\%	&	0.0040	&	&	2.79\%	&	0.0067	&	&	1.09 p.p.	&	&	4.65	&	0.77	\\
Lower bound	&	US\cent 7.3	&	High	&	&	70.80\%	&	0.0143	&	&	0.50\%	&	0.0023	&	&	0.70\%	&	0.0031	&	&	0.20 p.p.	&	&	10.51	&	2.22	\\
Cigarette	&	US\cent 11.5	&	High	&	&	70.10\%	&	0.0145	&	&	0.60\%	&	0.0024	&	&	0.85\%	&	0.0034	&	&	0.15 p.p.	&	&	6.90	&	0.89	\\
50\% decrease	&	US\cent 39.1	&	High	&	&	70.00\%	&	0.0150	&	&	0.40\%	&	0.0020	&	&	0.57\%	&	0.0029	&	&	0.17 p.p.	&	&	3.96	&	0.57	\\
25\% increase	&	US\cent 97.8	&	High	&	&	68.40\%	&	0.0147	&	&	0.30\%	&	0.0017	&	&	0.44\%	&	0.0025	&	&	0.14 p.p.	&	&	2.87	&	0.43	\\
\hline																																		\end{tabular}
				\begin{tablenotes}[para,flushleft]
	\footnotesize \textit{Notes}: Predicted measures for the representative man with access to marijuana. His level of education is high school (12 years of education), works, has a good self-perception of mental and physical health, no marijuana users in his network, consumes alcohol and cigarettes, lives in Medellín in a low socioeconomic stratum, and there is a drug dealer in her neighborhood. Pred. stands for prediction, and prob. is probability.\\	
	\textit{Source}: Authors' construction.
	\end{tablenotes}
	\end{minipage}}
\end{threeparttable}
%\end{table}
\end{sidewaystable}

\subsection{Tax revenues}

We perform some simulation exercises regarding the potential tax revenue that the government could collect from a tax on marijuana consumption under legalization. In particular, we use the posterior draws to simulate the model for all the individuals in the survey using the predictive framework of the previous subsection, assuming that the probability of access is equal to one for every one under a legal framework. We estimate the probability of consumption given access for each individual, and use it to sample from a Bernoulli distribution, if the realization of this experiment is 0, then the associated consumption is 0, if the realization is equal to 1, then, we predict their consumption conditional on access and use. Then, we use the expansion factors of the survey to estimate the potential yearly tax revenues. The survey represents 23.6 million individuals between 12 and 65 years-old in 2019, this is approximately 75\% of the total Colombian population in this age range. Thus, we should consider these predictions as underestimating the potential revenue. Although, the missing 25\% of the population is located in relatively isolated rural areas where potentially would have not legal marijuana suppliers. We also take into account that approximately 34\% of the demand of marijuana under a legal framework would be in the black market. This figure is based on the situation in the cigarettes market.\footnote{See \href{https://www.semana.com/salud/articulo/consumo-de-cigarrillos-34-de-cada-100-fueron-de-contrabando-esto-revelo-la-federacion-nacional-de-departamentos/202315/}{Estudio de incidencia del consumo de cigarrillos en colombia 2022.}} 

We set fourth potential scenarios where all of them assume that the difference between the average cost per gram (US\cent 1.33) and the price is equal to the tax. The first is a lower bound where the price is equal to the cost (US\cent 1.33) plus a tax that is equal to the cigarette tax (US\cent 5.9). The second scenario assumes that the price of marijuana is equal to the average price of a legal cigarette, the third assumes a price that is the 50\% of approximately the actual average price that a representative individual with access pays for a joint, and the fourth scenario uses a price that is 25\% more expensive than the latter. 

We can see in Table \ref{tab:tax} the results. We observe that the annually average tax revenue under legalization of marijuana in Colombia would be between USD 11.0 million and USD 54.2 million, depending on which taxation scheme is used. This is between 2.9\% to 14.4\% of the tax revenues of cigarettes, according to the Ministry of Finance and Public Credit, that reports tax revenues of cigarettes are equal to USD 374 million in 2019.\footnote{See \href{https://www.minhacienda.gov.co/webcenter/portal/Estadisticas}{Ministerio de hacienda y crédito público}} We should take into account that in Colombia, the amount of cigarettes per month of a smoker is 208, whereas the predicted average consumption of marijuana under the counterfactual of cigarette price is 57 per month, almost four times less. In addition, the predicted probability of use of marijuana is 2.5\% in this price setting under legalization, whereas the probability of cigarette is 9.5\% (reporting smoking in the last month), according to the ENCSPA survey, that is, approximately 4 times higher.

\begin{table}[h!]    
\centering \caption{Tax income experiments: Legalization of marijuana in Colombia.\label{tab:tax}}
	\begin{threeparttable}
		\resizebox{0.6\textwidth}{!}{\begin{minipage}{\textwidth}
				\begin{tabular}{l c c c}\hline
\hline							
\multirow{2}{*}{Price scenario}	 &	Price	&	Tax revenue	&	Tax income \\
 & per gram & per gram & USD million \\	
\hline							
Lower bound	&	US\cent 7.3	&	US\cent 6.0	&	USD 11.0	\\
Cigarette	&	US\cent 11.5	&	US\cent 10.2	&	USD 14.9	\\
50\% decrease	&	US\cent 39.1	&	US\cent 37.8	&	USD 32.0	\\
25\% increase	&	US\cent 97.8	&	US\cent 96.5	&	USD 54.2	\\
\hline							
						
\end{tabular}
				\begin{tablenotes}[para,flushleft]
	\footnotesize \textit{Notes}: Potential scenarios of tax income under different prices and taxes. This is yearly tax income in million USD using the average exchange rate in 2019 (COP/USD 3.274), taking into account that 34\% of marijuana demand would be in the black market, and using the expansion factors that account for 75\% of the Colombian population in 2019 between 12 and 65 years-old. The reference cost is US\cent 1.33.\\	
	\textit{Source}: Authors' construction.
	\end{tablenotes}
	\end{minipage}}
\end{threeparttable}
\end{table}

\section{Concluding remarks}\label{sec:Con}

We present an endogenous three-part model to estimate the demand of marijuana in Colombia that allows to infer the potential effects of its legalization finding heterogeneous effects among age groups and gender. Thus, we extend \cite{jacobi2016marijuana}'s proposal modeling simultaneously the three stages of marijuana demand (access, extensive and intensive margins), taking truncation into account.

The main estimation findings indicate that women have a lower probability of access, use, and quantity of consumption than men. Individuals over 30 also have a lower probability of access and use than younger individuals (20s and below), and individuals with good mental health also have a lower probability of access and use. Overall, we find that the demand for marijuana exhibits inelasticity (-0.45); moreover, there is no statistically significant difference in this elasticity in prices across age groups. These results are robust to different specifications, and access, price and intensive margin definitions.

We also find that a legalization policy would increase the probability of use from 2.3\% to 3.0\%, particularly affecting young individuals, from 4.3\% to 5.5\%, where risk perception is a relevant driver to curve marijuana demand. Therefore, under a legalization policy, marketing warning the potential bad consequences of consuming marijuana targeting younger individuals would be mandatory. Therefore, part of the potential tax revenues, which under a realistic setting would be approximately USD 32 million, should be invested in these warning campaigns. In addition, we have that given the relatively low price of marijuana in Colombia, this variable can be use to drive the intensive margin, rather than the extensive margin. In any case, warning campaigns should be complemented with taxes as a comprehensive strategy to mitigate negative effects associated with marijuana consumption.

We consider in this study the short-term effects of a legalization policy on the extensive and intensive margins of marijuana demand, and potential tax revenues from this activity. However, future research should consider long-term effects of this policy. In this point, the ``gateway hypothesis" is relevant as a legal framework for marijuana consumption may imply consequences on consumption of more toxic drugs due to the polydrug use. In addition, other aspects of a policy of legalization should be considered, for instance, effects on public health, labor and crime. The latter is particularly relevant in Colombia due to the relevance of marijuana in the micro traffic business.

 \clearpage
\bibliographystyle{apalike}
\bibliography{Referencias}
\clearpage

\clearpage

%\section*{Figures}

\appendix
\section{Variable construction}\label{sec:appendix1}
\subsection{Quantity consumed weighted by marijuana type}\label{quality}

There are different types of marijuana, among which the \textit{ENCSPA} highlights regular, corinto, creepy, and others. All of these have different intensities of active compounds (\textit{tetrahydrocannabinol} - THC). It is crucial to weigh the amounts consumed by the proportion of active compounds, thus converting the different varieties into one in common. Taking into account the studies of \cite{castano2017aportes} and \cite{pachon2012corinto}, we learn that the corinto and regular varieties cannot be considered statistically different in terms of their THC composition (therefore, for this study, they will be considered as the same variety); While the creepy variety can contain on average up to 4 times more THC than the regular one. With this in mind, the nearest neighbor algorithm is estimated to find the price of each variety for all individuals consuming marijuana. Taking advantage of the fact that there are individuals who only consume one class, a Euclidean distance is calculated from observable characteristics between individuals who consume more than one variety and individuals who only consume a single one, and the nearest neighbor price is associated. \\
Once we have the price for each marijuana type consumed by all marijuana consumers, we assume that the price reported in the survey corresponds to a weighted average of the consumption of each class; therefore, using the following equations, we can obtain the amounts per variety for creepy and regular:

%\vspace{5 mm}
%\begin{center}
    $$\text{Avg. price} = \frac{\text{price}_i\times \text{quantity}_i + \text{price}_j\times \text{quantity}_j}{\text{Total consumption}}$$
%\vspace{5 mm}
    $$\text{Total consumption} = \text{quantity}_i + \text{quantity}_j$$
%\end{center}
%\vspace{5 mm}   

Finally, after having the amounts per variety, the quantities of creepy are multiplied by 4 (since this is the proportion of more THC among the strains), leaving everything in terms of regular marijuana. As for users of the ``other" class (0.3\% of the total population and 7\% of marijuana users), it is not possible to know precisely the intensity of the active compound. Therefore, including them in the analysis is not possible, and they are not taken into account in this analysis.

\subsection{Prices}\label{prices}
    
We can find a direct measure of price by dividing the monthly expenditure by the average number of units consumed by individuals who report marijuana consumption in the last year. However, the price is not directly observed for individuals who report no consumption of marijuana. Thus, we follow the nearest neighbor algorithm to impute price: First, prices below 10\% and above 95\% of the specific drug price distribution are left out of consideration as we want to avoid a bias toward the imputed distribution due to extreme values. Second, we impute the price of marijuana for those individuals who report no marijuana consumption using the average price for the same municipality and stratum. Third, if there is not a marijuana consumer in the same municipality and stratum, we use the average price at the municipality level, then, the average price at the stratum level, and finally, the unconditional average price, in case there is not any match.

\subsection{Risks perceptions}\label{risk}
We created a single measure of risk perception from the questions asked in the survey about self-perceived risk according to the use of marijuana at different frequencies (rarely, sometimes, and frequently). Each of these three variables takes the ordered values of 1 if you consider it to be no risk, 2 slight risks, 3 moderate risks, and 4 high risks. To create a single measure we take the average between the three variables and divide it into three broad categories of low risk, medium risk, and finally high risk (independently of the frequency).

\subsection{Different access, prices, and quantities measures}\label{robustnessAppe}
We use different measures of access, prices, and quantities consumed of marijuana weighted by THC content to test the robustness of the estimates to small changes in the definitions of each variable from the survey data. In particular, we made variations in three key variables. The changes made are as follows.

%\begin{itemize}
%    \item Access: The access variable used in the main estimates is defined as 1 if the individual responds in the survey that it is easy for him/her to obtain marijuana and 0 if he/she responds that he/she would not know how to obtain it, it would be difficult or impossible. For the robustness exercise we relax the access measure somewhat by including within the individuals who have access those who report that it would be easy and difficult for them to get marijuana.

% \item Prices: The price variable used in the main estimates is achieved by dividing the expenditure by the quantity consumed. However, the ENCSPA 2019 survey additionally asks the question ``Do you know how much a marijuana cigarette or joint costs?" which in our view represents a more generalized measure of market prices with respect to the more individual-specific one that is inferred from their expenditures. 

% \item Quantities: Because the price variable is used to deduce by a nearest neighbor the possible price of each type of marijuana for all consumers, changing the price measure automatically necessitates using a quantity measure aligned with the new prices.
%\end{itemize}

Table \ref{tab:DescStatsRobust} shows the descriptive statistics of these three alternative measures compared to descriptive statistics in Table \ref{tab:DescStats}. We observe that the new definition of access implies overall more individuals having access to marijuana as expected. This increases 9 percentage points compared to the baseline setting. Note that columns 2 and 3 in Table \ref{tab:DescStatsRobust} correspond to the data filtered by the access variable under the main definition used in the primary analysis. The access variable under the definition used for heterogeneous effects is less strict. It differs from the original definition so that 22\% more of the people with access originally end up having access under this definition (Table \ref{tab:DescStatsRobust}, Column 2, Row 4). Regarding prices and quantities, we observe that the baseline and alternative settings are very similar. This means that these measures are both relatively consistent. 

%%% Descriptive statistics table 
\begin{table}[h!]\centering \caption{Summary statistics different measures of access, prices, and quantities.\label{tab:DescStatsRobust}}
	\begin{threeparttable}
		\resizebox{0.73\textwidth}{!}{\begin{minipage}{\textwidth}
\begin{tabular}{llccccccccc}
\hline
 & & All & \multicolumn{1}{l}{} & \multicolumn{3}{c}{Access (main) to marijuana} & \multicolumn{1}{l}{} & \multicolumn{3}{c}{Marijuana consumer} \\ \cline{3-3} \cline{5-7} \cline{9-11} 
 & & \multicolumn{1}{l}{} & \multicolumn{1}{l}{} & No & & Yes & & No & & Yes \\ \cline{5-5} \cline{7-7} \cline{9-9} \cline{11-11} 
Variable & & (1) & & (2) & & (3) & & (4) & & (5) \\ \cline{2-11} 
 & & & & & & & & & & \\
Access to marijuana & & \multicolumn{1}{l}{} & \multicolumn{1}{l}{} & \multicolumn{1}{l}{} & \multicolumn{1}{l}{} & \multicolumn{1}{l}{} & \multicolumn{1}{l}{} & \multicolumn{1}{l}{} & \multicolumn{1}{l}{} & \multicolumn{1}{l}{} \\
Main & & 0.58 & & 0.00 & & 1.00 & & 0.57 & & 1.00 \\
 & & (0.49) & & (0.00) & & (0.00) & & (0.5) & & (0.00) \\
Robustness & & 0.67 & & 0.22 & & 1.00 & & 0.66 & & 1.00 \\
 & & (0.47) & & (0.41) & & (0.00) & & (0.47) & & (0.00) \\
Quantity consumed & & & & & & & & & & \\
Main & & 1.01 & & 0.00 & & 1.75 & & 0.00 & & 43.08 \\
 & & (13.61) & & (0.00) & & (17.89) & & (0.00) & & (78.07) \\
Robustness & & 1.05 & & 0.00 & & 1.82 & & 0.00 & & 44.67 \\
 & & (14.04) & & (0.00) & & (18.44) & & (0.00) & & (80.36) \\
Price of marijuana & & & & & & & & & & \\
Main & & 0.83 & & 0.86 & & 0.80 & & 0.83 & & 0.84 \\
 & & (0.44) & & (0.48) & & (0.40) & & (0.43) & & (0.54) \\
Robustness & & 0.81 & & 0.84 & & 0.79 & & 0.81 & & 0.84 \\
 & & (0.3) & & (0.31) & & (0.28) & & (0.29) & & (0.37) \\
 & & & & & & & & & & \\
Sample size & & 49,414 & & 20,909 & & 28,505 & & 48,255 & & 1,159 \\ \hline
\end{tabular}

				\begin{tablenotes}[para,flushleft]
	\footnotesize \textit{Notes}: Standard deviations in parenthesis. This table presents descriptive statistics for ENCSPA 2019 regarding access and consumption of Marijuana in Colombia, as well as control variables. Column 1 shows the information for the entire sample. Columns 2 and 3 show the information for individuals without and with access to marijuana. Columns 4 and 5 show the information for not consumers and consumers, respectively. Individuals who are consumers should have access; this is internally consistent for more than 98\% of the population and inputted for the remaining 2\%. Prices are shown in 2019 USD and were converted using the average exchange rate in 2019 (3,274 USD/COP).\\	
	\textit{Source}: Authors' construction using ENCSPA data.
	\end{tablenotes}
	\end{minipage}}
\end{threeparttable}
\end{table}

\end{document}